\documentclass[12pt,a4paper]{article}

\usepackage{amsmath}
\usepackage{amssymb}
\usepackage{mathtools}
\usepackage{amsthm}
\usepackage[mathcal]{euscript}
\usepackage{bbm}
\usepackage{epsfig}

\usepackage{xcolor}
\usepackage[all]{xy}

\newcommand{\cref}{\ref}
\newcommand{\faktor}[2]{#1/#2}

\newcommand{\kh}{{\mc K}}
\newcommand{\rkh}{{\hat{\mc K}}}
\newcommand{\rh}{{\mc H}}
\newcommand{\rrh}{{\hat{\mc H}}}

\newcommand{\F}{\mathfrak{F}} 

\newcommand{\C}{\mathfrak{C}} 

\newcommand{\N}{\mathfrak{N}} 

\newcommand{\D}{\mathfrak{D}}

\renewcommand{\O}{\mathfrak{O}}

\newcommand{\R}{\mathfrak{R}}

\newcommand{\A}{\mathfrak{A}}
\renewcommand{\P}{\mc{P}}

\renewcommand{\S}{\mathfrak{S}}

\newcommand{\state}[1]{\S (#1)}

\newcommand{\B}{\ensuremath{\mathfrak{B}}}

\newcommand{\Nu}{\mathcal{V}}

\newcommand{\menge}[2]{\left\{ \, #1 \; \left| \; #2\,  \right. \right\}}

\newcommand{\fcond}[2]{#1 \; \forall \; #2}

\newcommand{\umenge}[3]{\menge{#1\,\in #2}{#3}}
\newcommand{\fumenge}[4]{\umenge{#1}{#2}{\fcond{#3}{#4}}}

\newcommand\restr[2]{{#1}{\raise-.5ex\hbox{\ensuremath|}_{#2}}}

\newcommand{\normcl}[1]{\left[#1\right]}

\newcommand{\bra}[1]{\left\langle \, #1 \, \right|}
\newcommand{\ket}[1]{\left| \, #1 \, \right\rangle}
\newcommand{\dirac}{\S_{\rm D}}

\theoremstyle{theorem}
\newtheorem{Theorem}{Theorem}[section]
\newtheorem{Proposition}[Theorem]{Proposition}
\newtheorem{Lemma}[Theorem]{Lemma}
\newtheorem{Corollary}[Theorem]{Corollary}
\newtheorem{Definition}[Theorem]{Definition}

\theoremstyle{remark}
\newtheorem{Remark}[Theorem]{Remark}
\newtheorem{Example}[Theorem]{Example}
\newcommand{\bco}{\begin{Corollary}}
\newcommand{\eco}{\end{Corollary}}

\newcommand{\bpr}{\begin{Proposition}}
\newcommand{\epr}{\end{Proposition}}

\newcommand{\btm}{\begin{Theorem}}
\newcommand{\etm}{\end{Theorem}}

\newcommand{\ben}{\begin{enumerate}}
\newcommand{\een}{\end{enumerate}}

\newcommand{\bit}{\begin{itemize}}
\newcommand{\eit}{\end{itemize}}

\newcommand{\bca}{\begin{cases}}
\newcommand{\eca}{\end{cases}}

\newcommand{\bre}{\begin{Remark}\rm}
\newcommand{\ere}{\end{Remark}}

\newcommand*{\bbm}{\begin{Remark}}
\newcommand*{\ebm}{\end{Remark}}

\newcommand{\ble}{\begin{Lemma}}
\newcommand{\ele}{\end{Lemma}}

\newcommand*{\bsz}{\begin{Proposition}}
\newcommand*{\esz}{\end{Proposition}}

\newcommand{\beq}{\begin{equation}}
\newcommand{\eeq}{\end{equation}}

\newcommand{\bbma}{\begin{bmatrix}}
\newcommand{\ebma}{\end{bmatrix}}

\newcommand{\bpma}{\begin{pmatrix}}
\newcommand{\epma}{\end{pmatrix}}

\newcommand*{\bbs}{\begin{Example}}
\newcommand*{\ebs}{\end{Example}}

\newcommand*{\bfg}{\begin{Corollary}}
\newcommand*{\efg}{\end{Corollary}}

\newcommand*{\bdf}{\begin{Definition}}
\newcommand*{\edf}{\end{Definition}}

\newcommand*{\bbw}{\begin{proof}}
\newcommand*{\ebw}{\end{proof}}

\newcommand*{\bpf}{\begin{proof}}
\newcommand*{\epf}{\end{proof}}

\newtagform{simple}{}{}
\newtagform{standard}{(}{)}
\newcommand{\eqqed}{\usetagform{simple}\tag{$\square$}}
\newcommand{\eqqedan}{\usetagform{simple}}
\newcommand{\eqqedaus}{\usetagform{standard}}

\allowdisplaybreaks

\newcommand{\II}{\mathbbm{1}}

\newcommand{\CC}{{\mathbb{C}}}

\newcommand{\RR}{{\mathbb{R}}}

\newcommand*{\res}{\upharpoonright}

\newcommand{\sitem}{\rm\item\it}

\newcommand{\SL}{{\mr{SL}}}

\newcommand{\SU}{{\mr{SU}}}
\newcommand{\su}{\mf{su}}

\newcommand{\ala}[1]{\begin{align*} #1 \end{align*}}

\DeclareMathOperator{\Aut}{\rm Aut}

\DeclareMathOperator{\Ad}{Ad}

\DeclareMathOperator{\im}{im}

\DeclareMathOperator{\tr}{tr}

\newcommand{\mc}[1]{\mathcal{#1}}
\newcommand{\mf}[1]{\mathfrak{#1}}
\newcommand{\mm}{\mu}
\newcommand{\mr}[1]{\mathrm{#1}}

\newcommand{\comment}[1]{}
\newcommand{\verweis}[1]{}
\newcommand{\todo}[1]{}
\renewcommand{\d}{{\mr d}}

\newcommand{\ddtn}{\left.\frac{\mr d}{\mr d t}\right|_{t=0}}

\newcommand{\cfg}{{\mc X}}
\newcommand{\pha}{{\mc P}}

\newcommand{\ve}{\varepsilon}

\newcommand{\vp}{\varphi}

\newcommand{\ctg}{\mr T^\ast}

\newcommand{\rref}[1]{{\rm \ref{#1}}}

\newcommand{\ol}[1]{\overline{#1}}
\newcommand{\ul}[1]{\underline{#1}}

\newcommand{\abs}{\hspace*{2.5mm}}
\newcommand*{\qeb}{\nopagebreak\hspace*{0.1em}\hspace*{\fill}{\mbox{\small$\blacklozenge$}}}

\newcommand{\prin}{1}
\newcommand{\sing}{0}

\begin{document}

\title{Stratified structure of the observable algebra \\of Hamiltonian lattice gauge theory}
\author{S.\ Knappe, G.\ Rudolph, M.\ Schmidt$^\ast$,\\ Institute for Theoretical Physics, University of Leipzig,\\ P.O. Box 100 920, D-4109 Leipzig, Germany.
\\
$^\ast$ Corresponding author}
\date{\today}
\maketitle

\begin{abstract}
We consider K\"ahler quantized models whose underlying classical phase space has a stratified structure induced from the Hamiltonian action of a compact Lie group. We show how to implement the classical stratification on the level of the $C^*$-algebra of observables and discuss the relation to the costratification 
(in the sense of Huebschmann) of the physical Hilbert space. Our analysis is based on 
the $T$-procedure as developed by Grundling and Hurst. We apply the general theory to Yang-Mills theory on a finite lattice, where the stratification is given by  the classical gauge orbit types. 
\end{abstract}

\newpage

\tableofcontents

%%%%%%%%%%%%%%%%%%%%%%%%%%%%%%%%%%%%%%%%%%%%%%%%%%%%%%%%%%%%%%%%%%%%%%%%%%%
%%%%%%%%%%%%%%%%%%%%%%%%%%%%%%%%%%%%%%%%%%%%%%%%%%%%%%%%%%%%%%%%%%%%%%%%%%%

\section{Introduction}

%%%%%%%%%%%%%%%%%%%%%%%%%%%%%%%%%%%%%%%%%%%%%%%%%%%%%%%%%%%%%%%%%%%%%%%%%%%
%%%%%%%%%%%%%%%%%%%%%%%%%%%%%%%%%%%%%%%%%%%%%%%%%%%%%%%%%%%%%%%%%%%%%%%%%%%

This paper is part of our research programme which aims at constructing Hamiltonian quantum gauge theory on a rigorous level. As we will explain below, it fills a certain gap 
in our understanding of the implementation of the classical gauge orbit type stratification at the quantum level. 

Our starting point is a finite-dimensional Hamiltonian lattice approximation of the gauge theory under consideration. On the classical level, this is a finite-dimensional Hamiltonian system with a symmetry. The corresponding quantum theory is obtained via canonical quantization. It is best described in the language of $C^*$-algebras with a field algebra which (for a pure gauge theory) 
may be identified with the algebra of compact operators on the Hilbert space of square-integrable functions over the product $G^N$ of a number of copies of the gauge group manifold $G$. Accordingly, the observable algebra is obtained via gauge symmetry reduction. We refer to \cite{qcd2, qcd3, qcd1, RS} for the study of this algebra, including its superselection structure. For the construction of the thermodynamical limit, see \cite{GR1,GR2}. 

If the gauge group is non-Abelian, then the action of the symmetry group in the corresponding classical Hamiltonian system necessarily has more than one orbit type. Accordingly, the reduced phase space obtained by symplectic reduction is a stratified symplectic space \cite{SjamaarLerman,OrtegaRatiu} rather than a symplectic manifold as in the case with one orbit type \cite{AbrahamMarsden}. The stratification is given by the orbit type strata. It consists of an open and dense principal stratum and several secondary strata. Each of these strata is invariant under the dynamics with respect to any invariant Hamiltonian. For case studies we refer to  \cite{cfg, cfgtop,FRS}. 

To study the influence of the classical orbit type stratification at the quantum level, we  use the concept of costratification of the quantum Hilbert space as developed by Huebsch\-mann \cite{Hue:Quantization}. A costratification is given by a family of closed subspaces, one for each stratum. Loosely speaking, the closed subspace associated with a certain classical stratum consists of the wave functions which are optimally localized at that stratum in the sense that they are orthogonal to all states vanishing at that stratum. The vanishing condition can be given sense in the framework of holomorphic quantization, where wave functions are true functions and not just classes of functions. In \cite{HRS} we have constructed this costratification for a toy model with gauge group $\SU(2)$ on a single lattice plaquette. As physical effects, we found a non-trivial overlap between distant strata and, for a certain range of the coupling, a very large transition probability between the ground state of the lattice Hamiltonian and one of the two secondary strata. In \cite{FJRS,FuRS,HoRS}, we made first steps towards extending this study to arbitrary finite lattices and arbitrary compact gauge group. \comment{In that case, there are non-trivial relations characterizing the classical gauge orbit strata which, in a first step, should be implemented at the quantum level. This problem has been solved in \cite{FuRS} using the above holomorphic picture.}

It has been unclear to us for many years how the above costratified structure manifests itself on the dual level, that is, on the level of the observable algebra. This problem is solved in the present paper. The key ingredient is the $T$-procedure for quantum systems with constraints as developed by Grundling and Hurst \cite{GH1, GH2}, combined with a fundamental theorem on hereditary $C^*$-subalgebras of $C^*$-algebras which is often referred to as the Open Projection Theorem, see \cite{Pedersen}. Using this tool, 
one finds that for every quantum system with constraints $(\mf F, \mf C)$ in the sense of Grundling and Hurst, there exists a projection in the strong closure of the algebra $\mf F$ in the universal representation such that the hereditary subalgebra $\mf D$, generated by the constraint set $\mf C$, is completely given by this projection. The arguments used apply to the 
general case of  K\"ahler quantized models whose underlying classical phase space has a stratified structure induced from the Hamiltonian action of a compact Lie group. Thus, 
in Section \ref{Str-QS} we deal with this general situation. We show how to implement the classical stratification on the level of the $C^*$-algebra of observables and discuss the relation to the costratification of the physical Hilbert space. Then, 
in Section \ref{Appl-HLGT}, we apply the general theory to Hamiltonian quantum gauge theory on a finite lattice.

%%%%%%%%%%%%%%%%%%%%%%%%%%%%%%%%%%%%%%%%%%%%%%%%%%%%%%%%%%%%%%%%%%%%%%%%%%%%%%%%
%%%%%%%%%%%%%%%%%%%%%%%%%%%%%%%%%%%%%%%%%%%%%%%%%%%%%%%%%%%%%%%%%%%%%%%%%%%%%%%%%%

\section{Background and motivation. An example}
\label{Background}
%%%%%%%%%%%%%%%%%%%%%%%%%%%%%%%%%%%%%%%%%%%%%%%%%%%%%%%%%%%%%%%%%%%%%%%%%%%%%%%%%
%%%%%%%%%%%%%%%%%%%%%%%%%%%%%%%%%%%%%%%%%%%%%%%%%%%%%%%%%%%%%%%%%%%%%%%%%%%%%%%%%

In the first part of this section, we explain the differential geometric background for the class of classical models we are interested in. We refer to \cite{Buch} for details. In the second part, we outline the canonical quantization procedure and the implementation of the classical gauge orbit stratification at the quantum level. All that is illustrated in terms of a toy model. In particular, 
we point out the physical significance of the gauge orbit stratification in quantum theory.

\subsection{The classical picture}

%%%%%%%%%%%%%%%%%%%%%%%%%%%%%%%%%%%%%%%%%%%%%

The mathematical model for a phase space endowed with a symmetry is a Hamiltonian $G$-manifold, that is, a triple $(M, \omega, G)$, where $(M,\omega)$ is a symplectic manifold and $G$ is a Lie group 
acting symplectically on $M$ and admitting a momentum mapping $\mu: M \to {\mf g}^\ast$, that is, a mapping from the phase space to the dual space of the Lie algebra $\mf g$ of $G$. Here, we are only interested in the case where $M$ and $G$ are finite-dimensional and $G$ is compact. The latter implies that the $G$-action is proper. If zero is a regular value of $\mu$ and $G$ acts freely on the level set $\mu^{-1}(0)$, then the reduced phase space 
$$
\hat M := \mu^{-1} (0)/ G
$$
is a symplectic manifold. This classical reduction is referred to as Marsden-Weinstein reduction \cite{AbrahamMarsden}. If the above regularity assumptions do not hold, then $\hat M$ 
is not a manifold, but a stratified symplectic space. That is, $\hat M$ is endowed with a smooth structure $C^\infty(\hat M)$ such that the following hold. 
\begin{enumerate}
\item Each stratum is a symplectic manifold. 
\item  $C^\infty(\hat M)$ carries the structure of a Poisson algebra. 
\item The embeddings of the strata into $\hat M$ are Poisson. 
\end{enumerate}
This fundamental result has been proven in \cite{SjamaarLerman}, see also \cite{OrtegaRatiu} for the extension to the case of non-zero level sets. For the general notion of a stratified space, see e.g. \cite{PflaumBook}. The strata of $\hat M$ are obtained as follows: for a subgroup $H \subset G$, let $M_{(H)}$ be the subset of all points whose stabilizer is conjugate to $H$, the subset of $M$ of orbit type $(H)$. The connected components of the orbit type subsets are called the strata of $M$. By standard arguments, the strata are embedded submanifolds and their union clearly coincides with $M$. Accordingly, one takes the orbit type subsets of $\hat M$ as
$$
\hat M_{(H)} = \left( M_{(H)} \cap \mu^{-1}(0) \right)/G \, . 
$$
The corresponding connected components constitute the strata of $\hat M$.
The orbit types carry a natural partial ordering given by reverse subconjugacy.  The smooth structure of $\hat M$ is provided by taking for $C^\infty (\hat M)$ the space $C^\infty(M)^G$  of smooth 
$G$-invariant functions on $M$. By $G$-invariance of $\omega$, $C^\infty(M)^G$ is a Poisson subalgebra of $C^\infty(M)$. This way, 
$C^\infty (\hat M)$ may be endowed with the structure of a Poisson space. Now, the proof of the Singular Reduction Theorem essentially rests on the local normal form of the momentum mapping as 
constructed by Marle and Guillemin and Sternberg. We refer to \cite{SjamaarLerman, OrtegaRatiu} for the study of the above stratified structure. In particular, it is important to note that 
it always contains a connected open and dense principal stratum. The remaining strata are referred to as secondary, or singular, or non-generic. Moreover, every stratum is invariant under the dynamics with respect to any invariant Hamiltonian.

As an example, let us consider the simplest non-trivial case provided by putting $N = 1$ in the lattice gauge model described in Subsection \ref{model}.
As explained in this subsection, the gauge symmetry reduction goes in two stages, first comes a harmless regular reduction implied by the choice of a lattice tree and thereafter one faces 
a singular reduction problem. The latter is discussed here for the case $N = 1$. After the regular reduction stage, the configuration space $Q$ coincides with the group manifold $G$ and 
the group of local gauge transformations is reduced to $G$. The action of $G$ on $Q$ is given given by conjugation. Thus, the partially reduced phase space is $\ctg G$ and 
the action of $G$ by conjugation on itself naturally lifts to a symplectic action on $\ctg G$. The lifted action admits the standard momentum mapping
$$
\mm : \ctg G \to \mf g^\ast
 \, , \quad
\mm(p)\big(X) := p(X_\ast)\,,
$$
where $p \in \ctg G$, $X \in \mf g$ and $X_\ast$ denotes the Killing vector field defined by $X$. A simple calculation shows that under the global trivialization
$ \ctg G \cong G\times \mf g$ induced by left-invariant vector fields and an invariant scalar product on $\mf g$, the lifted action is given by diagonal conjugation,
$$
g \cdot (a , A)
 = 
\big(g a g^{-1} , \Ad(g) A\big)
$$
and the associated momentum mapping is given by 
\beq
\label{G-ImpAbb-1}
\mu(a ,  A)
 =
 \Ad(a) A- A \,,
\eeq
see e.g.\ \cite[Section 10.7]{Buch}. The (fully) reduced phase space $\pha$ is obtained from $\ctg G$ by singular symplectic reduction at $\mm = 0$. This condition corresponds to the Gau{\ss} 
law constraint. 

\begin{figure}[h]

\unitlength1cm

\begin{center}

\begin{picture}(4,3.5)

\put(0,0.5){
 \put(0,0.27){
 \put(0,0){\put(0.05,0.1){\makebox(-0.1,-0.2)[tr]{$\pha_+$}}} 
 \put(4,0){\put(0.05,0.1){\makebox(-0.1,-0.2)[tl]{~~$\pha_-$}}} 
 \put(4,1.75){\put(0.05,0.1){\makebox(-0.1,-0.2)[tl]{~~$\pha_0$}}} 
 \put(2,-0.5){\put(0.05,0.1){\makebox(-0.1,-0.2)[tc]{$\pha = \big(T\times\mf t\big)/W$}}} 
 \put(0,-0.06){\circle*{0.2}}
 \put(4,-0.06){\circle*{0.2}}
 }
 \put(-2,0){
 \put(0,0.02){\epsfig{file=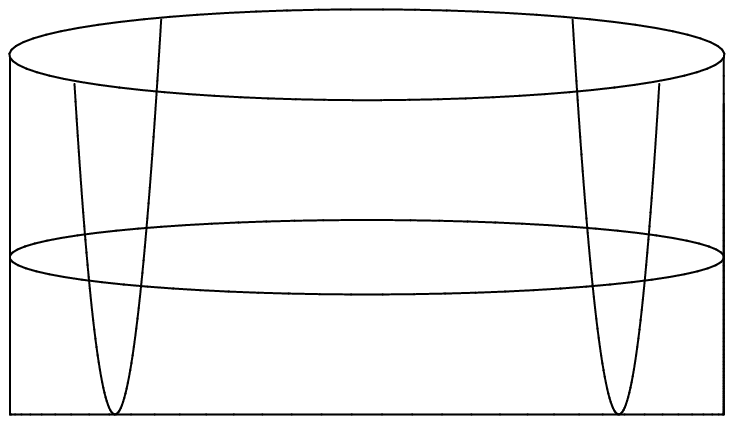,width=6cm,height=2cm}}
 }
}
 \end{picture}

\end{center}

\caption{\label{Abb-Kanu} 
Reduced phase space $\pha$}

\end{figure}

Let us determine the orbit type strata of $\pha$ for the special case $G = \SU(2)$. 
Let $Z$ denote the center of $G$ and let $T \subset G$ denote the subgroup of diagonal matrices. Clearly, $T$ is a maximal toral subgroup, isomorphic to $\mr U(1)$. Let $\mf t \subseteq \mf g$ be the Lie subalgebra associated with $T$. By \eqref{G-ImpAbb-1}, the vanishing of $\mu(a,A)$ implies that $a$ and $A$ commute. Hence, the pair $(a,A)$ is conjugate to an element of
$T\times \mf t$ and the injection $T\times\mf t \hookrightarrow G \times \mf g$ induces a homeomorphism of $\pha$ onto the quotient $(T\times\mf t)/W$, where $W$ is the Weyl group acting simultaneously on $T$ and $\mf t$ by permutation of the entries. If we identify $T$ with the complex unit circle and $\mf t$ with the imaginary axis,  then $W$
acts on $T$ by complex conjugation and on $\mf t$ by reflection. Hence, the reduced configuration space $\cfg \cong T/W$ is homeomorphic to a closed interval and the reduced phase space $\pha \cong (T\times\mf t)/W$ is homeomorphic to the well-known
canoe, see Figure \rref{Abb-Kanu}. 
Corresponding to the partitions $2=2$ and $2=1+1$, there are two orbit types, which we denote by $\sing$ and $\prin$, respectively. The orbit type subset $\cfg_\sing$ consists of the classes of $\pm\II$, i.~e., of the
endpoints of the interval; it decomposes into the connected components $\cfg_+$, consisting of the class of $\II$, and $\cfg_-$, consisting of the class of $-\II$. The orbit type subset $\cfg_\prin$ is connected and consists of the interior of the interval. 
The (secondary) orbit type subset $\pha_\sing$ consists of the classes of $(\pm\II,0)$, i.~e., of
the vertices of the canoe; it decomposes into the connected components $\pha_+$, consisting of the class of $(\II,0)$, and $\pha_-$, consisting of the class of $(-\II,0)$. The (principal) orbit type subset $\pha_\prin$ consists of the remaining classes, has dimension $2$ and is connected. It is worthwile to notice that the secondary strata constitute singularities of the natural Poisson 
structure on $\pha$, see \cite{HRS}.

%%%%%%%%%%%%%%%%%%%%%%%%%%%%%%%%%%%%%%

\subsection{The quantum picture}
\label{Q-picture}
%%%%%%%%%%%%%%%%%%%%%%%%%%%%%%%%%%%%%%%%%%%

The quantum theory is obtained via canonical quantization, see \cite{qcd2, qcd3, GR1, GR2}. The quantum reduction also goes in two stages. For the quantum reduction corresponding to the classical regular reduction stage we refer to \cite{qcd3}. By Theorem 5.2 of this paper, for the model at hand canonical quantization commutes with regular reduction. Thus, we may assume that the classical regular reduction is done and the canonical quantization starts at that stage, see Subsection \ref{CanQuant}. For the quantum theory so obtained the reduction of the residual gauge symmetry at the quantum level is performed (reduction after quantization). Finally, the quantum counterpart of the classical gauge orbit stratification is constructed.   
Let us illustrate all that for the case $N = 1$. The physical Hilbert space is 
$$
\rh :=  L^2(G) \, .
$$
It carries the generalized Schr\"odinger representation $\pi = (U,T)$,  where $T_f$ and $U_g$ are the multiplication and translation operators representing the gauge connection and the colour electric field, respectively.  The (residual) gauge transformation law at the quantum level is given by the following unitary representation of $G$ on $L^2(G)$: 
$$
(V_g \varphi)(h) := \varphi(g^{-1}\,h\, g) \, , \quad \varphi \in L^2(G)\, .
$$
Via $V$, we obtain the local gauge transformations of the quantum fields, see formulae \eqref{G-GTr-Tf-1} and \eqref{G-GTr-Tf-2}. 
Next, one observes that $\pi$ is a covariant representation of a certain $C^\ast$-dynamical system. The associated crossed product $C^\ast$-algebra coincides under 
$\pi$ with the algebra of compact operators ${\mathfrak K} \big( L^2(G) \big)$. The latter serves as the field algebra of the model.
Finally, we have a natural action by automorphism of the (residual) group of local gauge transformations $G$ on ${\mathfrak K} \big( L^2(G) \big)$. In the representation $\pi$ it is given by 
$g \to {\rm Ad}(V_g)$ via \eqref{G-GTr-Tf-1} and \eqref{G-GTr-Tf-2}. Thus, we can perform the reduction of the residual gauge symmetry at the quantum level. Under the representation $\pi$, this yields as the algebra of observables the algebra  of compact operators $  \mf K \big(\rrh\big)$ on the subspace 
$\rrh = \{ \Phi \in \rh \, \, | \, \, V_g ( \Phi) = \Phi \, \, \, \text{for all} \,  \, \, g \in G \} $ of $G$-invariant elements of $\rh$.

Now, we wish to construct a quantum counterpart of the classical gauge orbit stratification. For that purpose,  we need a notion of localization of quantum states on a given stratum. This can be accomplished by using the concept of costratification of the quantum Hilbert space as developed by Huebsch\-mann \cite{Hue:Quantization}, combined with a localization concept taken from the theory of 
coherent states. The latter only works  in the context of holomorphic (K\"ahler) quantization, where wave functions are true functions (and not just classes of functions). 
Thus, we have to limit our attention to cases, where the unreduced phase space is a (positive quantizable) K\"ahler manifold. The details will be presented in Subsection \ref{AlgObs-CostrHSp}, for the application 
to the lattice gauge model see Subsections \ref{Quantization-Reduction} and \ref{Strat-obsAlg}. We stress that, here, K\"ahler quantization and reduction with respect to the pointed gauge group also commute, see \cite[Thm.\ 5.3.1]{Woodward} and \cite{Hue:Quantization} for a general statement. Thus, like in the case of canonical quantization, we may restrict attention to the second stage, that is, reduction with respect to the residual gauge symmetry. Let us briefly outline the case $N = 1$, $G = \SU(2)$ as studied in \cite{HRS}. By composing the inverse polar mapping 
$$
G\times\mf g \to G_\CC
 \,, \quad 
(a,A) \mapsto a\exp(\mr i A)\, ,
$$
with the left global cotangent bundle trivialization, we obtain an equivariant diffeomorphism $\ctg G \to G_\CC$.
Via this diffeomorphism, the complex structure of $G_\CC$ and the symplectic structure of $\ctg G$ combine to a K\"ahler structure. Then, half-form K\"ahler quantization  on $G_\CC$ yields the Hilbert space 
$$
{\cal K} := HL^2(G_\CC, \mr d \nu_\hbar)
$$
of holomorphic functions on $G_\CC$ which are square-integrable with respect to the measure defined by \eqref{measure-nu}. Reduction then yields the closed subspace 
$$
\rkh := {\cal K}^G \equiv HL^2(G_\CC, \mr d \nu_\hbar)^G
$$
of $G$-invariants as the Hilbert space of the reduced system. In fact, $\rkh$ is the image of the canonical representation space 
$\rrh = L^2(G)^G$ of the observable algebra $  \mf K \big(\rrh\big)$ under the Segal-Bargmann transformation. This fact guarantees that all structures and statements obtained in the 
K\"ahler quantized context may be carried over to the standard canonical quantization context and vice versa. Being $G$-invariant, the elements of $\rkh$ are continuous functions on 
$\pha \cong (T\times\mf t)/W$.
Now, to construct the quantum counterpart of the classical stratification, to every stratum $\tau$ we associate the subspace $\rkh_\tau$ of elements of $\rkh$ which are 
localized on $\P_\tau$ in the following sense. Consider the vanishing subspace $\Nu_\tau$ of $\P_\tau$ in $\rkh$, defined by
$$
\Nu_\tau \coloneqq \umenge{\xi}{\rkh}{\restr{\xi}{\P_\tau}=0},
$$
and take $\rkh_\tau$ as its orthogonal complement in $\rkh$, that is
$$
\rkh=\Nu_\tau \oplus \rkh_\tau.
$$
By construction, the subspace associated with the principal stratum coincides with $\rkh$. The subspaces $\rkh_\pm$ associated with the secondary strata $\pha_{\pm}$ turn out to be one-dimensional. In terms of the orthonormal basis of normalized complex irreducible characters of $G_\CC = \SL(2,\CC)$, provided by the holomorphic Peter-Weyl Theorem and given by
$$
(\hbar\pi)^{-\frac{3}{4}}e^{-\hbar\beta(n+1)^2/2} \, \chi^\CC_n
\,,\quad
n=0,1,2,\dots
\, ,
$$
the subspaces $\rkh_\pm$ are spanned by the vectors
$$
\psi_\pm
 =
\sum_{n=0}^\infty (\pm 1)^n \, (n+1) \, \mr e^{-\hbar \beta^2 (n+1)^2}\chi^\CC_n
\,,
$$
respectively. See Remark \rref{Bem-SU2} for further details. Using this explicit description, we have found the following \cite{HRS}. 

\ben

\item There is an overlap between $\psi_+$ and $\psi_-$, which vanishes in the classical limit and which can be interpreted as a tunneling probability between the secondary strata.

\item The ground state of the Kogut-Susskind lattice Hamiltonian, which can be calculated rigorously for the model at hand, is localized almost completely at $\pha_+$. More precisely, the expectation value of the projector to $\rkh_+$ in that state is almost equal to $1$.

\een

The second point supports the conjecture that the gauge orbit stratification carries information about the 
spectral properties of the quantum Hamiltonian. The first case study (Schr\"odinger quantum mechanics on a double cone) pointing this out was provided by Emmerich 
and R\"omer \cite{emmroeme}. For a further discussion of the possible physical relevance of the singular gauge orbit strata we refer to section 8.1 in \cite{RS-Buch2}. Here, we only 
add that the singular strata lead to additional anomalies, see \cite{Scheck}, and that they can carry a magnetic charge. The impact of the latter fact was studied in the context of $4$-dimensional Yang-Mills theory with a $\theta$-term \cite{Asorey2} and of topological Chern-Simons theory \cite{Asorey1}, \cite{RSV} (leading to kinematical quantum nodes and providing arguments for quark confinement). 

It should be clear from the above discussion that we are {\bf not} presenting a new quantization procedure here. What we are constructing is an {\bf additional} structure of standard quantum theory taking into account the stratified structure of the classical phase space. On the level of quantum states, this additional structure (the costratification) was constructed already before \cite{HRS,FuRS,FJRS}. Here, we present its counterpart on the level of the observable algebra. Effectively, it results in some additional projection operators, which will allow us to address various issues on the level of observables. Of particular importance will be the study of the spectral properties of the (finite-lattice) quantum Hamiltonian. In \cite{FJRS} we already have constructed important tools, which will enable us to study the above mentioned projectors.

%%%%%%%%%%%%%%%%%%%%%%%%%%%%%%%%%%%%%%%%%%%%%%%%%%%%%%%%%%%%%%%%%%%%%%%%%%%
%%%%%%%%%%%%%%%%%%%%%%%%%%%%%%%%%%%%%%%%%%%%%%%%%%%%%%%%%%%%%%%%%%%%%%%%%%%

\section{Quantum Systems with Constraints}

%%%%%%%%%%%%%%%%%%%%%%%%%%%%%%%%%%%%%%%%%%%%%%%%%%%%%%%%%%%%%%%%%%%%%%%%%%%
%%%%%%%%%%%%%%%%%%%%%%%%%%%%%%%%%%%%%%%%%%%%%%%%%%%%%%%%%%%%%%%%%%%%%%%%%%%

%%%%%%%%%%%%%%%%%%%%%%%%%%%%%%%%%%%%%%%%%%%%%%%%%%%%%%%%%%%%%%%%%%%%%%%%%%%

\subsection[The $T$-procedure]{The \textit{\textbf{T}}-procedure}
\label{SectT-procedure}

%%%%%%%%%%%%%%%%%%%%%%%%%%%%%%%%%%%%%%%%%%%%%%%%%%%%%%%%%%%%%%%%%%%%%%%%%%%

Here, we present a summary of the $T$-procedure as developed by Grundling and Hurst, 
see  \cite{GH1, GH2,GL}. This procedure provides a way to implement quantum constraints 
in the language of $C^*$-algebras. Heuristically, a set of quantum constraints is given by a set ${\mf C} = \{C_i: i \in {\cal I} \}$ of operators $C_i$ on a Hilbert space $\mc H$ and the physical Hilbert space is then taken as  the subspace of $\mc H$ consisting of vectors which are annihilated by all the operators $C_i$. In this paper, we limit our attention to the case that all operators 
$C_i$ are bounded. Note that without loss of generality we may assume $\C = \C^*$. 

Below, the reader will see three subsequent assumptions. They ensure that the general procedure presented here produces meaningful physics.

\bdf
\label{defQSC}

A quantum system with constraints is a pair $(\F , \C )$ consisting of a unital 
C*-algebra $\F$, called the field algebra, and a self-adjoint subset $\C \subset \F$, called the constraint set. Given $(\F, \C)$, the physical state space is defined by
\begin{equation}
\label{eq:DiracConstrints}
\dirac \coloneqq\fumenge{\omega}{\state{\F}}{\pi_\omega (C) \Omega_\omega =0 \,\, }
{C \in \C}.
\end{equation}
Here $\state{\F}$ stands for the state space of $\F$ and 
$(\pi_\omega , \mc H_\omega , \Omega_\omega )$ denote the GNS-data of $\omega$. The elements of $\dirac$ are called Dirac states.

\edf
\noindent
{\bf Assumption 1:} All physical information is contained in the pair $(\F, \dirac)$.
\bigskip

Let $N_\omega := \left\{ F \in \F: \omega (F^* F) = 0 \right\}$ be the left kernel of 
$\omega \in \S(\F) $. Define 
$$
\N := \bigcap \left\{ N_\omega : \omega \in \dirac\right\} \, .
$$
By direct inspection, we have
$$
\dirac = \N^0 \cap \S(\F) \, ,
$$
where $\N^0$ denotes the annihilator of $\N$ in the dual of $\F$. By Lemma 3.13.5 
in \cite{Pedersen},  $\N$ coincides with the norm-closure $[\F \C]$ of the left ideal generated by \(\C\) in \(\F\). This yields the following equivalent characterization of 
the set of Dirac states. 

\bsz
\label{PropDiracStatesAnnihilatorIdeal}
Let $(\F , \C )$ be a quantum system with constraints. Then, 
$$
\dirac = [\F \C]^0 \cap \state{\F} \, .
$$
\esz

Next, recall that, as a closed left ideal, $\N$ defines a hereditary $C^*$-subalgebra $\D$ of $\F$ by
\beq
\label{D} 
\D :=  \N\cap\N^* \equiv  \normcl{\mathfrak{FC}} \cap \normcl{\mathfrak{CF}} \, .
\eeq
Moreover, the assignment
\beq
\label{Bij-LI-HA}
\N \mapsto \N\cap\N^* 
\eeq
defines a bijection from the set of closed left ideals of $\F$ onto the set of hereditary 
$C^*$-subalgebras of $\F$, see \cite[ 3.2.1]{Murphy}. From self-adjointness of $\C$ and from the fact that $\C$ is contained in $\N$, we obtain 
$$
\C \subset C^*(\C)\subset \D \, ,
$$
where $C^*(\C)$ denotes the $C^*$-algebra generated by $\C$.
The following theorem provides a criterion for the non-triviality of the $T$-procedure, see Theorem 2.2 in \cite{GL} for the proof. 

\btm
\label{DiracStates}

Let $(\F , \C )$ be a quantum system with constraints. Then, 

\ben

\sitem\label{TheoremT-procedureNontriviality}
\(\dirac \neq \emptyset \; \Leftrightarrow \; \mathbbm{1} \notin C^* (\C) \; \Leftrightarrow\; \mathbbm{1} \notin \D\),

\sitem\label{TheoremT-procedureDiracDAnnihilator}
\(\omega \in \dirac \; \Leftrightarrow \; \pi_\omega (\D) \Omega_\omega =0\).

\een

\etm

A constraint set $\C$ is called first class if $\mathbbm{1} \notin C^*(\C)$. Thus, if the set of constraints is first class, then the set of  Dirac states is non-trivial. 
Note that point 1 of  Theorem \ref{DiracStates} is equivalent to the assumption that $\N$ is a proper left ideal in $\F$.
\bigskip

\noindent
{\bf Assumption 2:} The constraint set is first class. 
\bigskip

Next, we define the $C^*$-subalgebra 
\beq
\label{O}
\O := \fumenge{F}{\F}{[F,D] := FD - DF \in \D\, \, }{D\in \D}.  
\eeq
Grundling and Hurst call $\O$ the weak commutant of the constraint set and observe that $\O$ is the $C^*$-algebraic counterpart of Dirac's observables.

\btm
\label{TheoremT-procedure}

Let \( (\F,\C) \) be a quantum system with constraints. Then the following hold.

\begin{enumerate}

\sitem\label{i-TheoremT-procedure-1}
$\D$ is the unique maximal C*-algebra in 
\(\bigcap \limits_{\omega \in \dirac} \ker \omega ,\)

\sitem\label{i-TheoremT-procedure-2}
$\O$ coincides with the restricted multiplier algebra of $\D$ in $\F$, that is,  
$$
\O = {\rm M}_\F (\D) = \left\{ F \in \F : FD \in \D, \,  DF \in \D \, \, \,  \forall \,
D \in \D  \right\},
$$

\sitem\label{TheoremT-procedureOWeakCom}
\(\O = \left\{ F \in \F : [F, \C] \subset \D \right\} \) ,

\sitem\label{DIdeal}
\(\D=\normcl{\mathfrak{OC}}=\normcl{\mathfrak{CO}} .\)

\end{enumerate}

\etm

For the proof we refer to \cite{GL}, Theorem 2.3.

\bbm

By point 3 of the above theorem, we have $\C' \subset \O$, where $\C'$  denotes the relative commutant of the constraint set in $\F$. Grundling and Hurst call $\C'$ the traditional observables. We will comment on the terminology in Example \ref{GaugeTh} below. By point 4, $\D$ is a closed two-sided ideal in 
$\O$, which is proper iff $\dirac \neq \varnothing$, that is, iff the constraints are first class.
\qeb

\ebm

The final step in the $T$-procedure consists in imposing the constraints described by $\D$. According to point \rref{i-TheoremT-procedure-2} of Theorem \rref{TheoremT-procedure}, $\mf D$ is an ideal in $\mf O$.

\bdf
\label{defR}

Let $(\F,\C)$ be a quantum system with constraints. The maximal C*-algebra respecting the constraints is defined as
\beq
\label{PhysOA}
\mathfrak{R} \coloneqq \faktor{\O}{\D} \, .
\eeq

\edf

Grundling and Hurst call $\R$ the algebra of physical observables.
\bigskip
 
\noindent
{\bf Assumption 3:} All physical information is contained in the pair $(\R, \S(\R))$.

\bbm
\label{T-procedure}

Here, we comment on the meaning of the above assumptions. To start with, Assumption 1 
simply says that given the field algebra $\F$ provided by the model under consideration, only those states may be considered as physically admissible that fulfil the quantum constraint. Concerning the heuristics behind the implementation of the constraints via state conditions as provided by Definition \ref{defQSC}, we have already made some remarks in the text above. This strategy clearly dates back to Dirac. 

Let us come to Assumption 2. A typical (quite generic \cite{Shan}) example of a classical second class constraint is provided by a pair of canonically conjugate momenta $(q,p)$. If we merely implement the constraints by putting $q | \psi\rangle = 0= p | \psi\rangle $, we obviously arrive at a contradiction: at one hand we obtain $\langle \psi| [q,p] | \psi \rangle = 0$ and on the other hand we have $[q,p] = i$. In particular, the latter tells us that $\mathbbm{1} \in C^* (\C)$. So, the above assumption that $\mathbbm{1} \notin C^* (\C)$ is needed to guarantee that we obtain a non-trivial set of Dirac states and the terminology ``first class'' on quantum level seems to be justified. (A direct adaptation of the notion of ``first class'' in the sense of Dirac to the quantum level has not been found up until now.) In \cite{GH2}, the authors have developed a procedure how to get rid of second class constraints on quantum level. Moreover, the reader may find a number of interesting examples there. 

Next, let us briefly discuss Assumption 3. In the study of concrete models it may, for instance, happen that the field algebra $\mf A$ is not unital, see our gauge model in Section \ref{Appl-HLGT}. Then, one needs to enlarge $\mf A$ by passing to a unital algebra $\mf F$ to get the $T$-procedure started. Clearly, thereafter one has to intersect the algebra $\mf R$ so obtained with $\mf A$ to obtain the correct algebra of observables. So, Assumption 3 is rather weak. It says that all physical information is contained in the pair $(\R, \S(\R))$, but it may well happen that $\mf R$ can and must be further reduced by physical arguments. 
\qeb

\ebm

%%%%%%%%%%%%%%%%%%%%%%%%%%%%%%%%%%%%%%%%%%%%%%%%%%%%%%%%%%%%%%%%%%%%%%%%%%%

\subsection{Open Projections}
\label{Openproj}

%%%%%%%%%%%%%%%%%%%%%%%%%%%%%%%%%%%%%%%%%%%%%%%%%%%%%%%%%%%%%%%%%%%%%%%%%%%

In this subsection, we discuss a useful tool which yields nice and more explicit 
presentations of the algebras introduced in the previous subsection. This was first observed in \cite{GL}. Recall that a projection in a $C^\ast$-algebra is an element $\mr q$ satisfying $\mr q^2=\mr q$ and $\mr q^\ast=\mr q$.
Thus, let there be given a constraint system $(\F, \C)$. Recall that $\D$ obtained by the $T$-procedure is a hereditary $C^*$-subalgebra of $\F$. This property will be basic in the sequel. It turns out that any hereditary $C^*$-subalgebra of a $C^*$-algebra $\mf B$ is determined by a unique projection in the strong closure of $\mf B$ in the universal representation. To understand this fact, start with the following simple observation: if  $\mr q$ is a projection in a $C^*$-algebra $\mf B$, then the $C^*$-subalgebra $\mr q \mf B \mr q$ is hereditary. Likewise, if $\mr q$ is a projection in a von Neumann algebra $\mf M$, then $\mr q \mf M \mr q$ is a strongly closed hereditary $C^*$-subalgebra of $\mf M$. For von Neumann algebras, the converse statement is also true, see Theorem 4.1.8 in \cite{Murphy} or Proposition 2.5.4 in \cite{Pedersen}. For the convenience of the reader, we include the proof.

\bsz
\label{PropProject}

Let $\mf M$ be a von Neumann algebra.  Then, the following hold.

\begin{enumerate}

\sitem\label{PropProjectHereditary}
For each strongly closed hereditary C*-subalgebra $\mf H$ of $\mf M$, there exists a unique projection $\mr q$ in $\mf M$ such that $\mf H = \mr q \mf M \mr q$. In fact, $\mr q$ is a unit in $\mf H$.

\sitem\label{PropProjectHereditary2} For each strongly closed left ideal $\mf J$ in $\mf M$ there exists a unique projection $\mr q$ in $\mf M$ such that $\mf J = \mf M \mr q$.

\end{enumerate}

\esz

\bpf

\rref{PropProjectHereditary}.\abs
Being strongly closed, $\mf H$ is a von Neumann algebra and hence possesses a unit $\mr q$. This unit is a projection in $\mf M$. Then, $\mf H = \mr q \mf H \mr q \subset \mr q \mf M \mr q$. Conversely, for any positive $a \in \mf M$, we have $a \leq \|a\| \II_{\mf M}$ and hence $\mr q a \mr q \leq \|a\| {\mr q} \in \mf H$. Since $\mf H$ is hereditary, then $\mr q a \mr q \in \mf H$. It follows that $\mr q \mf M \mr q \subset \mf H$. To prove uniqueness of $\mr q$, let $\mr q \in \mf M$ be a projection such that $\mf H = \mr q \mf M \mr q$. Being von Neumann, $\mf M$ has a unit. Hence, $\mr q \in \mf H$. Since $\mr q (\mr q a \mr q) = \mr q a \mr q = (\mr q a \mr q) \mr q$, the projection $\mr q$ is a unit in $\mf H$. It is thus unique.

\rref{PropProjectHereditary2}.\abs
$\mf J \cap \mf J^\ast$ is a strongly closed hereditary $C^\ast$-subalgebra of $\mf M$. Hence, by point \rref{PropProjectHereditary}, there exists a unique projection $\mr q$ in $\mf M$ such that $\mf J \cap \mf J^\ast = \mr q \mf M \mr q$. Then, $\mf J = \mf M (\mf J \cap \mf J^\ast) = \mf M \mr q \mf M \mr q$. Now, $\mf M \mr q \mf M \mr q \subset \mf M \mr q$ and, since $\mf M$ has a unit, $\mf M \mr q \mf M \mr q \supset \mf M \mr q$.
\epf

This proposition can be extended to arbitrary hereditary $C^\ast$-subalgebras of $C^*$-algebras as follows. Let $\pi^u : \mf B \to \mf B(\mc H^u)$ be the universal representation of $\mf B$ \cite[Sec.\ 3.7]{Pedersen}. Since $\pi^u$ is faithful, we may identify $\mf B$ with its image under $\pi^u$. Let $\mf B''$ be the enveloping von Neumann algebra, ie.\ the double commutant of $\mf B$ in $\mf B(\mc H^u)$. Thus, $\mf B \subset \mf B''$. The subsequent theorem is part of a set of statements in \cite[3.11.10]{Pedersen}, see also Appendix 1 in \cite{GL}.

\btm[Open Projection Theorem]
\label{ThmOpenProject}

Let $\mf H$ be a hereditary $C^*$-sub\-algebra of a $C^*$-algebra $\mf B$ in $\B({\cal H})$. Then, there exists a projection $\mr q^u$ in $\mf B''\subset\mf B(\mc H^u)$ such that
\beq
\label{OpenProjProp}
\mf H = (\mr q^u \mf B'' \mr q^u) \cap \mf B \, .
\eeq

\etm

We take up the following notion from \cite[3.11.10]{Pedersen}.

\bdf
\label{OpenProj}

Let ${\mf B}$ be a $C^*$-algebra. A projection $\mr q^u$ in $\mf B''$ is called open if $\mr q^u$ belongs to the strong closure in $\mf B''$ of the hereditary $C^*$-subalgebra $(\mr q^u \mf B'' \mr q^u) \cap \mf B $ of $\mf B$.

\edf

\bbm

By the above discussion, the open projections are in bijection with the hereditary
$C^*$-subalgebras of ${\mf B}$ via the mapping
$$
\mr q^u \to (\mr q^u \mf B'' \mr q^u) \cap {\mf B} \, .
$$
Analogously, by point 2 of Proposition \ref{PropProject}, the open projections are in bijection with the closed left ideals via the mapping
$$
\mr q^u \to (\mf B'' \mr q^u) \cap {\mf B}
$$
and, by \cite[3.11.10]{Pedersen}, with the weak *-closed faces containing $0$ of the quasistate space ${\rm Q} ({\mf B})$ via the mapping
\beq
\mr q^u \to \left\{ \omega \in {\rm Q}({\mf B}): \omega(\mr q^u) = 0 \right\} \, .
\eeq
For the latter, recall that every state $\omega \in \state{\F}$ has a unique continuation onto $\F''$, which we denote by the same symbol.
\qeb

\ebm

Applying Theorem \ref{ThmOpenProject} to a quantum system with constraints, we obtain the following.

\bfg
\label{F-D-uvlDst}

For every quantum system with constraints $(\F,\C )$ there exists a unique projection $\mr q^u$ in $\F''$ such that 
$$
\D = (\mr q^u \F'' \mr q^u)\cap\F = (\mr q^u \F \mr q^u)\cap\F\,.
$$

\efg

Analogously, the algebras $ \O$ and $\R$ as well as the set of Dirac states $\dirac$ may be characterized in terms of $\mr q^u$, see Appendix 1 in \cite{GL}.

\bsz
\label{S-DOR-uvlDst}

Let \((\F,\C)\) be a quantum system with constraints and let $\mr q^u$ be the projection provided by Corollary \rref{F-D-uvlDst}. Then, the following hold.

\ben

\sitem 
\label{i-S-DOR-uvlDst-Dir}
$\dirac = \{\omega \in \state{\F} | \omega(\mr q^u)=0\}$,

\sitem
\label{i-S-DOR-uvlDst-O}
$
\O
 = 
\{a \in \F | \mr q^u a(\II - \mr q^u) = 0 = (\II - \mr q^u) a \mr q^u\}
 = 
(\mr q^u)' \cap \F
$,

\sitem 
\label{i-S-DOR-uvlDst-R}
$\mf R \cong (\II - \mr q^u)\big((\mr q^u)' \cap \F\big)$.

\een

Here, $(\mr q^u)'$ denotes the commutant of $\mr q^u$ in $\F''$.

\esz

Consider the orthogonal decomposition of the universal Hilbert space $\mc H^u$:
\beq
\label{Dec-Hu}
{\cal H}^u = \mr q^u {\cal H}^u \oplus (\II - \mr q^u) {\cal H}^u \, .
\eeq

\bfg
\label{F-DOR-uvlDst}

Let $(\F,\C)$ be a quantum system with constraints. Relative to the decomposition \eqref{Dec-Hu}, we have the following presentations of the $C^*$-algebras $\D$, $\O$  and $\R$. 
\begin{align*}
\D
&= 
\left\{
F \in \F 
\left|
F = \bbma D & 0 \\ 0 & 0 \ebma
\,,~
D \in \mr q^u \F \mr q^u
\right.\right\}
\,,
\\
\O
&=
\left\{
F \in \F
\left|
F = \bbma A & 0 \\ 0 & B \ebma
\,,~
A \in \mr q^u \F \mr q^u
\,,~
B \in (\II-\mr q^u) \F (\II-q^u)
\right.\right\}
\\
\mf R
& \cong
\left\{
\tilde F \in (\II-\mr q^u) \F
\left|
\tilde F = \bbma 0 & 0 \\ 0 & R \ebma
\,,~
R \in (\II-\mr q^u) \F (\II-\mr q^u)
\right.\right\}
\subset \mf F''
\,.
\end{align*}

\efg

\bbw

The block presentation for $\D$ follows directly from Corollary \ref{F-D-uvlDst}. To find that of $\O$, according to Proposition \ref{S-DOR-uvlDst}, we have to intersect the commutant of $\mr q^u$ in $\F''$ with $\F$. This yields the subalgebra of $\F$ consisting of the elements of the form
$$
F = \bbma A & 0 \\ 0 & B \ebma
\,,
$$
where $A \in \mr q^u \F'' \mr q^u$ and $B \in (\II-\mr q^u) \F'' (\II-q^u)$. Since $F \in \F$, we have in fact
$$
A = \mr q^u F \mr q^u \in \mr q^u \F \mr q^u
\,,\qquad
B = (\II-\mr q^u) F (\II-q^u) \in (\II-\mr q^u) \F (\II-q^u)
\,.
$$
Finally, according to point \ref{i-S-DOR-uvlDst-R} of Proposition \ref{S-DOR-uvlDst}, to obtain a subalgebra of $\mc F''$ isomorphic to $\mf R$, we have to multiply the block representation of $\O$ from the left by $(\II-q^u)$. This yields the assertion.
\ebw

\bbm

Consider the block presentation of $\O$ given in Corollary \rref{F-DOR-uvlDst}. While for the elements of $(\mr q^u)' \subset \F''$, the blocks $A$ and $B$ can be chosen independently from one another, the requirement that $F = A + B$ be in $\F$ puts them into correlation to one another, so that in the block presentation of $\O$, they can {\em not} be chosen independently.
\qeb

\ebm

Next, we consider a faithful cyclic representation $(\pi, \cal H)$ of $\F$. Let $\tilde\pi$ denote the strong operator continuous extension of $\pi$ from $\F$ to $\F''$. By III.5.2.10 in \cite{Blackadar}, $\tilde\pi$ is faithful, too. Define
\beq\label{G-D-q}
\mr q := \tilde\pi(\mr q^u)
\,,\qquad
\mr p := (\II_{\mc H} - \mr q)
\,.
\eeq
We obtain the following orthogonal decomposition of the Hilbert space $\cal H$:
\beq
\label{Dec-H}
{\cal H} = \mr q {\cal H} \oplus \mr p {\cal H} \, .
\eeq

\bsz
\label{S-DOR-Dst}

Let $(\F,\C)$ be a quantum system with constraints and let $(\pi, \cal H)$ be a faithful cyclic representation of $\F$. Then, the following hold.

\ben

\sitem 
\label{i-S-DOR-Dst-D}
$\pi(\D) = (\mr q \, \pi(\F) \, \mr q) \cap \pi(\F)$,

\sitem
\label{i-S-DOR-Dst-O}
$\pi(\O) = \mr q' \cap \pi(\F)$,

\sitem 
\label{i-S-DOR-Dst-R}
$\mf R \cong \mr p \big(\mr q' \cap \pi(\F)\big)$.

\een

Here, $\mr q'$ denotes the commutant of $\mr q$ in $\B(\mc H)$.

\esz

\bbw

\rref{i-S-DOR-Dst-D}.\abs 
According to Corollary \rref{F-D-uvlDst}, $\pi(\D) = \tilde\pi(\D) = \tilde\pi\big((\mr q^u \F \mr q^u) \cap \F\big)$. By injectivity of $\tilde\pi$, this equals $\tilde\pi(\mr q^u \F \mr q^u) \cap \tilde\pi(\F)$. Since $\tilde\pi$ is an algebra homomorphism, this yields the assertion.

\rref{i-S-DOR-Dst-O}.\abs 
According to point \rref{i-S-DOR-uvlDst-O} of Proposition \rref{S-DOR-uvlDst}, and by injectivity of $\tilde\pi$, 
$$
\pi(\O)
= 
\tilde\pi(\O) 
= 
\tilde\pi\big((\mr q^u)' \cap \F\big)
= 
\tilde\pi\big((\mr q^u)'\big) \cap \pi(\F)
\,. 
$$
If $A \in (\mr q^u)'$, then $[\tilde\pi(A),\mr q] = \tilde\pi([A,\mr q^u]) = 0$. This shows that $\tilde\pi\big((\mr q^u)'\big) \subset \mr q'$ and hence $\pi(\O) \subset \mr q' \cap \pi(\F)$. Conversely, if $F \in \F$ so that $\pi(F) \in \mr q'$, then $\tilde\pi([F,\mr q^u]) = [\pi(F),\mr q] = 0$. Then, injectivity of $\tilde\pi$ implies $F \in (\mr q^u)'$ and hence $\pi(F) \in \pi(\O)$. Consequently, $\mr q' \cap \pi(\F) \subset \pi(\O)$.

\rref{i-S-DOR-Dst-R}.\abs 
Since $\tilde\pi$ is injective, composition of the isomorphism of point \rref{i-S-DOR-uvlDst-R} of Proposition \rref{S-DOR-uvlDst} with $\tilde\pi$ yields an isomorphism of $\mf R$ with $\tilde\pi\big((\II-\mr q^u) \big((\mr q^u)' \cap \F\big)\big)$. By the homomorphism property of $\tilde\pi$ and point \rref{i-S-DOR-Dst-O}, this yields the assertion.
\ebw

The following is now proved by analogy with Corollary \rref{F-DOR-uvlDst}, with $\F''$ replaced by $\B(\mc H)$.

\bfg
\label{F-DOR-Dst}

Let $(\F,\C)$ be a quantum system with constraints and let $(\pi, \cal H)$ be a faithful cyclic representation of $\F$. Then, relative to the decomposition \eqref{Dec-H}, we have the following presentations of the $C^*$-algebras $\D$, $\O$  and $\R$. 
\ala{
\pi(\D)
&= 
\left\{
f \in \pi(\F)
\left|
f = \bbma d & 0 \\ 0 & 0 \ebma
\,,~
d \in \mr q \pi(\F) \mr q
\right.\right\}
\,,
\\
\pi(\O)
&=
\left\{
f \in \pi(\F)
\left|
f = \bbma a & 0 \\ 0 & b \ebma
\,,~
a \in \mr q \pi(\F) \mr q
\,,~
b \in \mr p \pi(\F) \mr p
\right.\right\}
\,,
\\
\mathfrak{R}
& \cong
\left\{
\tilde f \in \mr p \pi(\F)
\left|
\tilde f = \bbma 0 & 0 \\ 0 & r \ebma
\,,~
r \in \mr p \pi(\F) \mr p
\right.\right\}
\,.
}
\qed

\efg

\bbm\label{Bem-D-tipi}

While it is convenient and practical to identify $\F$ with its image under the universal representation, at times it may be puzzling, in particular when further identifications enter the game. Let us distinguish between them here to give a clear definition of the extension $\tilde\pi$ to $\F''$ of the representation $\pi$ of $\F$. Since $\pi^u$ is faithful, there exists a unique $*$-homomorphism $\hat\pi : \pi^u(\F) \to \B(\mc H)$ satisfying $\hat\pi \circ \pi^u = \pi$. It is $\hat\pi$ which defines the representation $\tilde\pi$ of $\F''$ via extension. That is, we have the following commuting diagram.
$$
\xymatrix{
{\pi^u(\F)} \ar[rr]^{\subset} \ar[rrd]^{\hat\pi} 
& & 
{\F''} \ar[d]^{\tilde\pi} 
\\
{\F} \ar[u]_{\pi^u}  \ar[rr]^{\pi} 
& & 
{\B(\mc H)}
} 
$$

\ebm

\bbs[Observable algebra of lattice gauge theory]
\label{GaugeTh}

In the forthcoming subsection \ref{CanQuant}, the field and observable algebras of Hamiltonian quantum gauge theory on a finite lattice is discussed. In \cite{GR1} it has been shown that the observable algebra of this model may be obtained via the $T$-procedure. Let us briefly recall this application. As explained in Subsection \ref{CanQuant}, for a given gauge group $G$, under the generalized Schr\"odinger representation, the field algebra $\A_0$ of the model is given as the algebra of compact operators ${\mathfrak K} (\cal H)$ on the Hilbert space ${\cal H} \cong  L^2(G^N)$, see formulae \eqref{G-quHilbSp} and \eqref{G-FieldAlg} below. Clearly, $\A_0$ is not unital. So, to get the $T$-procedure started, we extend $\A_0$ to the unital crossed product $C^*$-algebra
$$
\F := ( \A_0 \oplus \CC) \rtimes_\alpha {\cal G}_d \, ,
$$
where ${\cal G}_d$ is the group of local gauge transformations (a product of copies of $G$ over all lattice sites), endowed with the discrete topology, and $\alpha$ represents the (strongly continuous) action of $\cal G$ on $\A_0$. Next, consider the unitary constraint $\C = U_{{\cal G}_d} - \II$, where 
$U_{{\cal G}_d}$ denotes the unitary representation of the group of local gauge transformations on $\cal H$. Starting from the pair $(\F, \C)$ so defined and using the above open projection method, one obtains 
that $\R$ defined by  \eqref{PhysOA} coincides with the algebra of observables 
${\mathfrak K} ({\cal H}^G)$ given by Theorem \rref{T-ObsAlg-Rep} below, which is denoted by $\mf A$ there. 

Note the slight difference in terminology: in the above cited original papers 
\cite{qcd2, qcd3} and in the notation used in this subsection, $\O$ is called the algebra of gauge invariant elements and $\R$ is referred to as the observable algebra.    
\qeb

\ebs

In the subsequent sections, we will see another application of the $T$-procedure in lattice gauge theory.

%%%%%%%%%%%%%%%%%%%%%%%%%%%%%%%%%%%%%%%%%%%%%%%%%%%%%%%%%%%%%%%%%%%%%%%%%%%
%%%%%%%%%%%%%%%%%%%%%%%%%%%%%%%%%%%%%%%%%%%%%%%%%%%%%%%%%%%%%%%%%%%%%%%%%%%

\section{Stratified Quantum Systems}
\label{Str-QS}

%%%%%%%%%%%%%%%%%%%%%%%%%%%%%%%%%%%%%%%%%%%%%%%%%%%%%%%%%%%%%%%%%%%%%%%%%%%
%%%%%%%%%%%%%%%%%%%%%%%%%%%%%%%%%%%%%%%%%%%%%%%%%%%%%%%%%%%%%%%%%%%%%%%%%%%

In this section we consider a physical system with finite-dimensional classical phase space $\P$. We assume that $\P$ has a non-trivial stratified structure. A typical example of that type is the reduced phase space of Hamiltonian quantum gauge theory on a finite lattice as will be  discussed in Section \ref{Appl-HLGT}. We implement the classical stratification at the quantum level by using the general concept of costratification of the quantum Hilbert space as developed by Huebsch\-mann \cite{Hue:Quantization}. In \cite{HRS} we have constructed such a costratification for a toy model of Hamiltonian lattice gauge theory with gauge group $\SU(2)$ by combining the abstract concept of costratification  with a localization concept for quantum systems taken from the theory of coherent states. This localization concept only works in the context of holomorphic (K\"ahler) quantization, where wave functions are true functions (and not just classes of functions). Thus, in this section we will always assume that the unreduced phase space is a (positive quantizable) K\"ahler manifold. 

Up until now, it has been unclear how to implement the stratified structure on the level of the observable algebra of the quantum system. This problem will be solved below. We  will construct a stratified structure on the observable algebra which, in a sense, is dual to the costratification of the physical Hilbert space.

%%%%%%%%%%%%%%%%%%%%%%%%%%%%%%%%%%%%%%%%%%%%%%%%%%%%%%%%%%%%%%%%%%%%%%%%%%%

\subsection{Costratified Hilbert Space}
\label{AlgObs-CostrHSp}

%%%%%%%%%%%%%%%%%%%%%%%%%%%%%%%%%%%%%%%%%%%%%%%%%%%%%%%%%%%%%%%%%%%%%%%%%%%

Let us start with the definition of the abstract notion of costratification 
\cite{Hue:Quantization}. Let $N$ be a stratified space and let ${\cal C}_N$ be the 
category whose objects are the strata of $N$ and whose morphisms are the inclusions 
$Y' \subset \overline{Y}$, where $Y$ and $Y'$ are strata such that $Y' \cap \ol Y \neq \varnothing$.

\bdf

A costratified Hilbert space relative to $N$ is a contravariant functor from ${\mc C}_N$
to the category of Hilbert spaces, with bounded linear maps as morphisms. 

\edf

This means that a costratified Hilbert space  relative to $N$ assigns a Hilbert space 
${\cal H}_Y$ to every stratum $Y$, together with a bounded linear map 
${\cal H}_{Y_2} \to {\cal H}_{Y_1}$ for each inclusion $Y_1 \subset \overline{Y_2}$.
Whenever $Y_1 \subset \overline{Y_2}$ and $Y_2 \subset \overline{Y_3}$, the composition of 
${\cal H}_{Y_3} \to {\cal H}_{Y_2}$ with ${\cal H}_{Y_2} \to {\cal H}_{Y_1}$ coincides 
with the bounded linear map ${\cal H}_{Y_3} \to {\cal H}_{Y_1}$ associated with the inclusion $Y_1 \subset \overline{Y_3}$.

Let us apply this concept to K\"ahler quantization with symmetries. Thus, assume that we are given a (positive quantizable) K\"ahler manifold $(M,J,\omega)$ endowed with the Hamiltonian action of a compact Lie group $G$ and let $\P$ be the stratified symplectic space obtained by zero level symplectic reduction. That is, $\P$ decomposes into disjoint symplectic manifolds,
$$
\P=\bigcup\limits_{\tau\in\mathrm{T}} \P_\tau
$$
fulfilling the frontier condition:
\beq
\label{eq:DefFrontierCond}
\P_\tau \cap \ol{\P_{\tau'}} \neq \varnothing  \quad \text{implies} \quad 
\P_\tau \subset \ol{\P_{\tau'}}\, ,
\eeq
for all $\tau,\tau'\in\mathrm{T}$. Note that the stratification carries a natural partial ordering defined by 
$$
\tau \leq \tau' :\Leftrightarrow \P_\tau \subset \ol{\P_{\tau'}} 
\,.
$$
We will loosely refer to $\mr T$ as the set of strata. 

For simplicity, in what follows, we restrict attention to the situation where the complex Hermitian line bundle obtained by K\"ahler quantization is trivial. Then, the Hilbert space $\kh$ consists of holomorphic functions on $M$ which are square-integrable wrt.\ a certain measure obtained by the quantization procedure. An example of that type is provided by the lattice gauge model discussed in Section \ref{Appl-HLGT}, see \eqref{measure-nu} for the measure. We define
$$
\rkh := \kh^G
$$
to be the subspace of $G$-invariant functions. We assume that the elements of $\rkh$ can be identified with functions on $\pha$. This assumption is fulfilled in Hamiltonian lattice gauge theory. These functions are then continuous. 

As explained in Subsection \ref{Q-picture} for the example discussed there, to construct a costratified Hilbert space for $\P$, to each $\tau\in \mathrm{T}$ we associate the subspace $\rkh_\tau$ of $\rkh$ of elements which are localized on $\P_\tau$ in the sense that they are orthogonal to the vanishing subspace 
$$
\Nu_\tau \coloneqq \umenge{\xi}{\rkh}{\restr{\xi}{\P_\tau}=0}
$$
of $\P_\tau$ in $\rkh$, ie.,
$$
\rkh=\Nu_\tau \oplus \rkh_\tau.
$$
By construction, if $\P_\tau \subseteq \ol{\P_{\tau'}}$, then $\Nu_\tau\supseteq \Nu_{\tau'}$ by continuity and, thus, $\rkh_\tau\subseteq \rkh_{\tau'}$. The system $\big\{ \rkh_\tau \big\}$, together with the orthogonal projections $\rkh_{\tau_2} \to \rkh_{\tau_1}$ whenever $\P_\tau \subseteq \ol{\P_{\tau'}}$, constitute a costratified Hilbert space relative to $\P$. As the construction is based on the chosen Hilbert space $\rkh$, we call the above structure a costratification of $\rkh$ relative to $\P$.

\bbm~

\ben

\item In physical applications, $M$ is usually given by the cotangent bundle $\ctg Q$ of some configuration space $Q$. For example, this is the case in Hamiltonian lattice gauge theory, to be discussed below. For that model, there is an obvious K\"ahler structure on $\ctg Q$ induced by the polar mapping of the gauge group, see \eqref{G-poldec-1}. However, combining embedding arguments taken from real algebraic geometry \cite{FSS} and appropriate deformation arguments, one can diffeomorphically identify any $\ctg Q$ with a neighbourhood of a real algebraic variety in its complexification. Via this diffeomorphism, one obtains a K\"ahler structure on $\ctg Q$. Clearly, the symplectic structure underlying this K\"ahler structure may differ from the canonical cotangent bundle symplectic structure. This argument is taken from a contribution by Tim Perutz to an online discussion, see \cite{Perutz}. Moreover, there are many special situations, where a K\"ahler structure of $\ctg Q$ may be constructed directly, see e.g.\ \cite{Mykytyuk} for the case where $Q$ is a rank one symmetric space of the compact type and \cite{Szoeke} for the case where $Q$ is a compact real analytic Riemannian manifold.

\item As we have seen above, K\"ahler quantization is useful for implementing the classical stratification on the level of pure states. On the other hand, it has the drawback that most of the relevant quantum observables cannot be obtained by this method. Therefore, it is desirable to have a Segal-Bargmann transformation between the Hilbert spaces obtained by K\"ahler quantization and by canonical quantization. In the case of Hamiltonian lattice gauge theory, a Segal-Bargmann transformation exists, because the unreduced phase space is given by the cotangent bundle of a Lie group and, thus, the theory developed by Hall applies \cite{Hall:SBT,Hall:cptype}, see Remark \rref{Segal-Bargmann}. This theory has been generalized to the case of symmetric spaces of both compact \cite{Stenzel} and noncompact complex type \cite{HallMitchell}.
\qeb

\een

\ebm

%%%%%%%%%%%%%%%%%%%%%%%%%%%%%%%%%%%%%%%%%%%%%%%%%%%%%%%%%%%%%%%%%%%%%%%%%%%

\subsection{Stratified Observable Algebra}
\label{str-AlgObs}

%%%%%%%%%%%%%%%%%%%%%%%%%%%%%%%%%%%%%%%%%%%%%%%%%%%%%%%%%%%%%%%%%%%%%%%%%%%

In this subsection, we show how to implement the stratification of the classical reduced phase space on the level of quantum observables. For that purpose, we introduce the following $C^\ast$-algebraic counterpart of the notion of costratified Hilbert space.

\bdf

A stratified $C^\ast$-algebra relative to the stratified space $N$ is a covariant functor from $\mc C_N$ to the category of $C^\ast$-algebras, with $\ast$-morphisms as morphisms.

\edf

This means that a stratified $C^\ast$-algebra relative to $N$ assigns a $C^\ast$-algebra $\mf A_Y$ to every stratum $Y$, together with a $C^\ast$-morphism $\mf A_{Y_1} \to \mf A_{Y_2}$ for each inclusion $Y_1 \subset \ol{Y_2}$. Whenever $Y_1 \subset \ol{Y_2}$ and $Y_2 \subset \ol{Y_3}$, the composition of $\mf A_{Y_1} \to \mf A_{Y_2}$ with $\mf A_{Y_2} \to \mf A_{Y_3}$ coincides with the morphism $\mf A_{Y_1} \to \mf A_{Y_3}$ associated with the inclusion $Y_1 \subset \ol{Y_3}$.

While the definition of stratified $C^\ast$-algebra (the observables) given here seems natural to us in that it is dual to the definition of costratified Hilbert space (the pure states), at the present stage it is just a working definition. Clearly, to the notion of stratified $C^\ast$-algebra there should correspond a notion of costratification for the corresponding full state space, and not only for the subset of pure states. This might require a sharpening of the conditions to be imposed on a stratified $C^\ast$-algebra. As this is work in progress, we do not discuss it here.

Let $\mf A$ be the field algebra of the quantum system under consideration, realized as a $C^\ast$-subalgebra of $\B(\rkh)$. In general, $\mf A$ need not be unital, like in the case of the gauge model studied in Section \rref{Appl-HLGT}. To construct from $\mf A$ a stratified $C^\ast$-algebra of observables relative to $\pha$, we assume that there exists a dense linear subspace $\mc R \subset \rkh$ which forms an algebra under pointwise multiplication of functions and which satisfies the following two natural conditions.

\ben

\item For every stratum $\pha_\tau$, the vanishing ideal of $\pha_\tau$ in $\mc R$ is dense in the vanishing subspace $\mc V_\tau$ of $\pha_\tau$ in $\rkh$ (the denseness condition).

\item For every stratum $\pha_\tau$, there exist finitely many relations $r_{\tau,1} , \dots , r_{\tau,n_\tau} \in \mc R$ generating the vanishing ideal of $\pha_\tau$ in $\mc R$. That is,

\ben

\item[2.1] $\ol{\pha_\tau} = \{x \in \pha : r_{\tau,i}(x) = 0 ~ \forall \, i\}$ (the zero locus condition).

\item[2.2] If $\psi \in \mc R$ vanishes on $\pha_\tau$, then 
$\psi = \sum_i \psi_i r_{\tau,i}$ for some $\psi_i \in \rkh$ (the radical ideal condition).

\een

\een

The denseness condition means that $\mc R$ contains all the information about the stratification that is carried by $\rkh$. Condition 2 ensures that the (closures of) strata can be treated as zero loci in the sense of algebraic geometry. In Hamiltonian lattice gauge theory, both conditions are satisfied, see Subsection \rref{Strat-obsAlg}. Here, the unreduced phase space is a product of copies of the complexified gauge group $G_\CC$ and thus it is an affine variety acted upon by the reductive algebraic group $G_\CC$. Its coordinate ring is given by the corresponding ring of representative functions. Accordingly, $\mc R$ may be chosen to be the subring of invariant representative functions. It is likely that the above conditions hold in more general situations, for example in some cases of affine K\"ahler manifolds acted upon by reductive algebraic groups. Another class of interest is provided by compact quantizable K\"ahler manifolds which are in fact projective algebraic manifolds by Kodaira embedding, see eg.\ \cite{Schlichenmaier}. Their symplectic quotients correspond to categorical quotients in the sense of geometric invariant theory \cite{GuiSte,Hue:Quantization,KempfNess,Kirwan,Mumford}. 

In the sequel, we apply the $T$-procedure. We take 
$$
\mf F := \B(\rkh)
\,.
$$
To define the constraint set, we choose an orthonormal basis $\{\psi_\alpha : \alpha \in A\}$ spanning $\mc R$ by finite linear combinations and put
$$
\mr q_{\tau,i,\alpha}
 := 
 \frac{
\ket{r_{\tau,i} \, \psi_\alpha} \bra{r_{\tau,i} \, \psi_\alpha}
 }{
\|{r_{\tau,i} \, \psi_\alpha}\|^2
 }
\,,\qquad
i = 1 , \dots, n_\tau
 \,;~
\alpha \in A
\,.
$$
These are the orthogonal projections onto the one-dimensional subspaces spanned by the functions $r_{\tau,i} \, \psi_\alpha$. We define
$$
\mf C_\tau
 := 
 \{
\mr q_{\tau,i,\alpha}
 :
i = 1 , \dots, n_\tau
 \,;~
\alpha \in A
 \}
\,.
$$
Clearly, $\mf C_\tau$ is self-adjoint. Application of the $T$-procedure to the quantum system with constraints given by $(\mf F,\mf C_\tau)$ yields a hereditary $C^\ast$-subalgebra $\mf D_\tau$, its weak commutant $\mf O_\tau$ and the corresponding algebra of physical observables $\mf R_\tau$. By Corollary \rref{F-D-uvlDst}, there exists a unique projection $\mr q_\tau^u$ in $\F''$ so that $\mf D_\tau = (\mr q_\tau^u \F \mr q_\tau^u) \cap \F$. Consider the identical representation $\pi$ of $\F = \B(\rkh)$ on $\rkh$ and let $\mr q_\tau$ and $\mr p_\tau$ be the complementary projections on $\rkh$ defined via $\pi$ by \eqref{G-D-q}. Since $\pi$ is faithful and cyclic, the algebras $\D_\tau$, $\O_\tau$ and $\mf R_\tau$ are presented by the formulae in Corollary \rref{F-DOR-Dst}, where $\pi$ can be omitted. Since $\F = \B(\rkh)$, we have $\mr q_\tau \in \F$ and hence $\mr p_\tau \in \F$. Since $\mr q_\tau \in \F$, we have
\beq\label{G-Dt}
\mf D_\tau = \mr q_\tau \mf F \mr q_\tau
\,.
\eeq
In particular, $\mf D_\tau$ is strongly closed. Then, Proposition \rref{PropProject} yields that $\mr q_\tau$ is a unit in $\mf D_\tau$. Since $\mr p_\tau \in \F$, $\mf R_\tau$ may be identified with a subalgebra of $\mf F$. Using this identification, we define
$$
\mf A_\tau := \mf R_\tau \cap \mf A \equiv (\mr p_\tau \mf F \mr p_\tau) \cap \mf A
\,.
$$
Note that, in general, the projectors $\mr q_\tau$ and $\mr p_\tau$ need not belong to $\mf A_\tau$.

\bbm\label{Bem-D-tipi-idDst}

Writing down the definition of $\mr q_\tau$ under the identification of $\F$ with its image under the universal representation $\pi^u$ might be puzzling, because, in effect, this produces the equation $\mr q_\tau = \mr q_\tau^u$ which does not seem to make sense, because it equates operators on different spaces. However, this comes as no surprise, because in the special situation at hand, if we identify $\F$ with its image under $\pi^u$, we do identify operators on $\rkh$ with operators on the universal Hilbert space $\mc H^u$. Thus, the equation we get just tells us that $\mr q_\tau$ is the operator on $\rkh$ which corresponds to the operator $\mr q^u_\tau$ on $\mc H^u$ under the identification of $\F$ with $\pi^u(\F)$, and so it makes perfect sense. The situation can be unravelled also formally along the lines of Remark \rref{Bem-D-tipi}. In the notation introduced there, we have $\pi^u(\F) = \F''$ and hence $\tilde\pi = \hat\pi$, so that the defining equation for $\mr q^u$ reads $\mr q^u = \hat\pi(\mr q_\tau^u)$. In addition, since $\pi$ is the identical mapping, $\hat\pi$ inverts $\pi^u$, and the diagram given in Remark \rref{Bem-D-tipi} shrinks to 
$$
\xymatrix{
\B(\rkh) = \F ~~ \ar@<0.5ex>[rr]^{\pi^u} 
& & 
~~ \pi^u(\F) \subset \B(\mc H^u) \ar@<0.5ex>[ll]^{\hat\pi}
}
$$

\ebm

\bsz\label{S-SOA}~

\ben

\sitem\label{i-S-SOA-subset}
If $\pha_\tau \subset \ol{\pha_{\tau'}}$, then $\mf A_\tau \subset \mf A_{\tau'}$. 

\sitem\label{i-S-SOA-SOA}
The system $\{\mf A_\tau\}$, together with the natural inclusion mappings $\mf A_\tau \to \mf A_{\tau'}$ whenever $\pha_\tau \subset \ol{\pha_{\tau'}}$, constitute a stratified $C^\ast$-algebra relative to $\pha$. 

\een

\esz

\bpf

It suffices to prove point \rref{i-S-SOA-subset}. If $\pha_\tau \subset \ol{\pha_{\tau'}}$, by continuity, the zero locus condition yields $(r_{\tau',i'})_{\res\pha_\tau} = 0$ for all $i' = 1 , \dots , n_{\tau'}$. Then, the radical ideal condition implies that $r_{\tau',i'} = \sum_i \vp_i \, r_{\tau,i}$ with $\vp_i \in \mc R$. Expanding
$$
\vp_i \, \psi_\alpha = \sum_\beta C_{i,\alpha,\beta} \psi_\beta
\,,
$$
we obtain
 \ala{
\mr q_{\tau',i',\alpha}
 & =
 \frac{
\ket{\sum_{i,\beta} C_{i,\alpha,\beta} \, r_{\tau,i} \, \psi_\beta}
\bra{\sum_{j,\gamma} C_{j,\alpha,\gamma} \, r_{\tau,j} \, \psi_\gamma}
 }{
\|{r_{\tau',i'} \, \psi_\alpha}\|^2
 }
\\
 & =
\sum_{i,j,\beta,\gamma}
 \frac{
C_{i,\alpha,\beta} \, C_{j,\alpha,\gamma}
 }{
\|{r_{\tau',i'} \, \psi_\alpha}\|^2
 }
 ~~
\mr q_{\tau,i,\beta}
 \big(
\ket{r_{\tau,i} \psi_\beta} \! \bra{r_{\tau,j} \psi_\gamma}
 \big)
\mr q_{\tau,j,\gamma}
\,.
 } 
This shows that $\mr q_{\tau',i',\alpha} \in \mf D_\tau$ for all $i',\alpha$. Hence, $\mf D_{\tau'} \subset \mf D_\tau$. Since $\mr q_\tau$ is a unit in $\mf D_\tau$,
$$
\mr q_\tau \mr q_{\tau'} = \mr q_{\tau'} = \mr q_{\tau'} \mr q_\tau 
\,.
$$
This implies 
$$
\mr p_\tau \mr p_{\tau'} = \mr p_\tau = \mr p_{\tau'} \mr p_\tau
$$
and hence 
$
\mr p_\tau \mf F \mr p_\tau
 =
\mr p_{\tau'} \mr p_\tau \mf F \mr p_\tau \mr p_{\tau'}
 \subset
\mr p_{\tau'} \mf F \mr p_{\tau'}
\,.
$
\epf

Next, we relate the projection $\mr p_\tau$ to the closed subspace $\rkh_\tau$ associated with $\pha_\tau$.

\bsz\label{S-SOA-KHR}

For every $\tau \in \mr T$, we have $\mr q_\tau\rkh = \Nu_\tau$ and $\mr p_\tau \rkh = \rkh_\tau$.

\esz

Thus, the subalgebra $\mf A_\tau$ associated with the stratum $\pha_\tau$ is given by the same projection as the closed subspace $\rkh_\tau$ associated with $\pha_\tau$. First, this implies that $\mf A_\tau$ is independent of the choice of the basis used in the construction. Second, in view of Corollary \rref{F-DOR-Dst}, this implies that $\mf A_\tau$ may be identified with a $C^\ast$-subalgebra of $\B(\rkh_\tau)$.

\bpf

It suffices to show $\mr q_\tau \rkh = \Nu_\tau$.

($\subset$)\abs Let $\xi \in \rkh$ be given. We have to show that $(\mr q_\tau \xi)_{\res \pha_\tau} = 0$. Since $\mr q_\tau \in \mf D_\tau$, and since $\mf D_\tau$ is contained in the norm-closed left ideal in $\mf F$ generated by $\mf C_\tau$, $\mr q_\tau$ is the norm-limit of some sequence $(a_k)$ in $\mf F$ whose members are of the form 
$$
a_k = \sum_{i,\alpha} \mr q_{\tau,i,\alpha} \, a_{k,i,\alpha}
$$
with $a_{k,i,\alpha} \in \mf F$ and with $a_{k,i,\alpha} \neq 0$ for only finitely many $\alpha$. Since $(r_{\tau,i} \, \psi_\alpha)_{\res\pha_\tau} = 0$ for all $i,\alpha$, we have $(a_k \xi)_{\res \pha_\tau} = 0$ for all $k$. Since $a_k \to \mr q_\tau$ in norm, $a_k \xi \to \mr q_\tau \xi$ in the $L^2$-norm. Since for holomorphic $L^2$-functions, $L^2$-norm convergence implies pointwise convergence, see e.g.\ Lemma 1.4.1 in \cite{Krantz}, we obtain $(\mr q_\tau \xi)_{\res \pha_\tau} = 0$, as was to be shown.

($\supset$)\abs In view of the denseness condition, it suffices to show that $\vp \in \mr q_\tau(\rkh)$ for every $\vp \in \mc R$ vanishing on $\pha_\tau$. Given such $\vp$, the radical ideal condition implies that $\vp = \sum_i \vp_i r_{\tau_i}$ for some $\vp_i \in \rkh$. Expanding $\vp_i = \sum_\alpha C_{i,\alpha} \psi_\alpha$, we see that 
\beq\label{G-SOA-KHR-1}
\vp
 = 
\sum_{i,\alpha} C_{i,\alpha} \, r_{\tau,i} \, \psi_\alpha
 = 
\sum_{i,\alpha} C_{i,\alpha} \, \mr q_{\tau,i,\alpha}(r_{\tau,i} \psi_\alpha)
\,.
\eeq
This shows that $\vp$ belongs to the closed subspace spanned by the images of the projections $\mr q_{\tau,i,\alpha}$. Since the latter belong to $\mf D_\tau$, Corollary \rref{F-DOR-Dst} implies that their images are contained in $\mr q_\tau(\rkh)$.
\epf

We introduce the following terminology.

\bdf

We say that an operator $a \in \B(\rkh)$ is 

\ben

\sitem annihilating for $\pha_\tau$ if $\im(a) \subset \Nu_\tau$,

\sitem localized on $\pha_\tau$ if $\ker(a) \supset \Nu_\tau$.

\een

In addition, we say that a set of annihilating operators for $\pha_\tau$ is generating if the hereditary $C^\ast$-subalgebra of $\B(\rkh)$ generated by this set contains all one-dimensional projections which project onto subspaces spanned by elements of $\mc R$ vanishing on $\pha_\tau$. 

\edf

\bsz\label{F-ann}

For every $\tau \in \mr T$, the following holds.

\ben

\sitem\label{i-S-ann-D} 
The elements of $\mf D_\tau$ are annihilating for $\pha_\tau$.

\sitem\label{i-S-ann-R} 
The elements of $\mf R_\tau$, and thus in particular the elements of $\mf A_\tau$, are localized on $\pha_\tau$.

\sitem\label{i-S-ann-C} 
The constraint set $\mf C_\tau$ is generating. 

\een

\esz

\bpf

Points \rref{i-S-ann-D} and \rref{i-S-ann-R} follow instantly from Proposition \rref{S-SOA-KHR} and Corollary \rref{F-DOR-Dst}. To prove point \rref{i-S-ann-C}, let $\mr q$ be a one-dimensional projection onto a subspace spanned by some $\vp \in \mc R$ such that $\vp_{\res\pha_\tau} = 0$. Assuming $\vp$ to be normalized, we have $\mr q = \ket\vp \bra\vp$. By an argument given in the proof of Proposition \rref{S-SOA-KHR}, $\vp$ can be written in the form \eqref{G-SOA-KHR-1}. Then,
 \ala{
\mr q
 & = 
\sum_{i,j;\alpha,\beta} C_{i,\alpha} \, \ol{C_{j,\beta}}
 \, 
\ket{r_{\tau,i} \psi_\alpha} \! \bra{r_{\tau,j} \psi_\beta}
\\
 & = 
\sum_{i,j;\alpha,\beta} C_{i,\alpha} \, \ol{C_{j,\beta}}
 ~
\mr q_{\tau,i,\alpha} 
\big(\ket{r_{\tau,i} \psi_\alpha} \! \bra{r_{\tau,j} \psi_\beta}\big) 
\mr q_{\tau,j,\beta} 
 }
and thus $\mr q \in \mf D_\tau$.
\epf

In the $T$-procedure, instead of $\mf C_\tau$, we may take any countable generating set of annihilating operators for $\pha_\tau$.

\bsz\label{S-SOA-ann}

Let $\tau \in \mr T$. If $\mf C \subset \mf F$ is a countable generating set of operators which are annihilating for $\pha_\tau$, then the $T$-procedure applied to the quantum system with constraints $(\mf F,\mf C)$ yields $\mf D_\tau$, $\mf O_\tau$ and $\mf R_\tau$.

\esz

As a consequence, the observable algebra $\mf A_\tau$ associated with the stratum $\pha_\tau$ can be obtained from any countable generating set of annihilating operators for $\pha_\tau$.

\bpf

Let $\mf D$ denote the hereditary $C^\ast$-subalgebra of $\mf F$ generated by $\mf C$. By the same argument as for $\mf D_\tau$, we find a unique projection $\mr q$ in $\mf F$ such that $\mf D = \mr q \mf F \mr q$. In particular, $\mf D$ is strongly closed, and Proposition \rref{PropProject} yields that $\mr q$ is a unit in $\mf D$. It suffices to show that $\mr q = \mr q_\tau$. For that purpose, in view of Proposition \rref{S-SOA-KHR}, it suffices to show that $\mr q \rkh = \Nu_\tau$. 

($\subset$)\abs This follows by the same argument as in the proof of Proposition \rref{S-SOA-KHR}. 

($\supset$)\abs We show that $\big((\II - \mr q) \rkh\big) \cap \Nu_\tau = \{0\}$. Assume, on the contrary, that there exists $\xi \in \rkh$ such that 
\beq\label{G-S-SOA-ann-1}
(\II - \mr q) \xi \neq 0
\,,\qquad
((\II - \mr q) \xi)_{\res\pha_\tau} = 0
\,.
\eeq
Denote $\vp := (\II - \mr q) \xi$. By the denseness condition, $\vp$ is the limit of a sequence $(\vp_n)$ in $\mc R$ whose members vanish on $\pha_\tau$. We may assume $\vp$ and all $\vp_n$ to be normalized. Then, the sequence of projections $\big(\ket{\vp_n}\!\bra{\vp_n}\big)$ converges in norm to $\ket{\vp} \bra{\vp}$. Since $\ket{\vp_n} \bra{\vp_n}$ is annihilating for $\pha_\tau$, by the generating property of $\mf C$, we have $\ket{\vp_n} \bra{\vp_n} \in \mf D$ for all $n$. Since $\mf D$ is closed, then $\ket\vp\bra\vp \in \mf D$. Since $\mr q$ is a unit in $\mf D$, we have $\mr q \ket{\vp} \bra{\vp} = \ket{\vp} \bra{\vp}$. On the other hand, $\mr q \vp = 0$ and hence $\mr q \ket{\vp} \bra{\vp} = 0$ (contradiction). Hence, an element $\xi \in \rkh$ satisfying \eqref{G-S-SOA-ann-1} does not exist.
\epf

%%%%%%%%%%%%%%%%%%%%%%%%%%%%%%%%%%%%%%%%%%%%%%%%%%%%%%%%%%%%%%%%%%%%%%%%%%%
%%%%%%%%%%%%%%%%%%%%%%%%%%%%%%%%%%%%%%%%%%%%%%%%%%%%%%%%%%%%%%%%%%%%%%%%%%%

\section{Application to Hamiltonian lattice gauge theory}
\label{Appl-HLGT}

%%%%%%%%%%%%%%%%%%%%%%%%%%%%%%%%%%%%%%%%%%%%%%%%%%%%%%%%%%%%%%%%%%%%%%%%%%%
%%%%%%%%%%%%%%%%%%%%%%%%%%%%%%%%%%%%%%%%%%%%%%%%%%%%%%%%%%%%%%%%%%%%%%%%%%%

\subsection{The model}
\label{model}

%%%%%%%%%%%%%%%%%%%%%%%%%%%%%%%%%%%%%%%%%%%%%%%%%%%%%%%%%%%%%%%%%%%%%%%%%%%
%%%%%%%%%%%%%%%%%%%%%%%%%%%%%%%%%%%%%%%%%%%%%%%%%%%%%%%%%%%%%%%%%%%%%%%%%%%

Let $G$ be a compact Lie group and let $\mf g$ be its Lie algebra. Later on, we will specify $G = \SU(2)$, but for the time being, this is not necessary. Let $\Lambda$ be a finite spatial lattice and let $\Lambda^0$, $\Lambda^1$ and $\Lambda^2$ denote, respectively, the sets of lattice sites, lattice links and lattice plaquettes. For the links and plaquettes, let there be chosen an arbitrary orientation. In lattice gauge theory with gauge group $G$ in the Hamiltonian approach, gauge fields (the variables) are approximated by their parallel transporters along links and gauge transformations (the symmetries) are approximated by their values at the lattice sites. Thus, the classical configuration space is the space $G^{\Lambda^1}$ of mappings $\Lambda^1 \to G$, the classical symmetry group is the group $G^{\Lambda^0}$ of mappings $\Lambda^0 \to G$ with pointwise multiplication and the action of $g \in G^{\Lambda^0}$ on $a \in G^{\Lambda^1}$ is given by  
\beq
\label{G-Wir-voll}
(g \cdot a)(\ell) := g(x) a(\ell) g(y)^{-1}\,,
\eeq
where $\ell \in \Lambda^1$ and $x$, $y$ denote the starting point and the endpoint of $\ell$, respectively. The classical phase space is given by the associated Hamiltonian $G$-manifold \cite{AbrahamMarsden,Buch} and the reduced classical phase space is obtained from that by symplectic reduction \cite{OrtegaRatiu,Buch,SjamaarLerman}. We do not need the details here. 
Dynamics is ruled by the classical counterpart of the Kogut-Susskind lattice Hamiltonian, see \cite{KS}. When identifying $\ctg G$ with $G \times \mf g$, and thus $\ctg G^{\Lambda^1}$ with $G^{\Lambda^1} \times \mf g^{\Lambda^1}$, by means of left-invariant vector fields, the classical Hamiltonian is given by 
\beq
\label{Hamiltonian-C}
H(a,E)
 = 
\frac{\lambda^2}{2 \delta} \sum_{\ell \in \Lambda^1}^N \Vert E(\ell) \Vert^2
 -
\frac{1}{\lambda^2 \delta} \sum_{p \in \Lambda^2} \left(\tr a(p) + \ol{\tr a(p)}\right)
 \,,
\eeq
where $a \in G^{\Lambda^1}$, $\lambda$ denotes the coupling constant, $\delta$ denotes the lattice spacing and $a(p)$ denotes the product of $a(\ell)$ along the boundary of the plaquette $p \in \Lambda^2$ in the induced orientation. The trace is taken in some chosen unitary representation. Unitarity ensures that the  Hamiltonian does not depend on the choice of plaquette orientations. Finally, $\Lambda^1 \ni \ell \mapsto E(\ell) \in \mf g^{\Lambda^1}$ is the classical colour electric field (canonically conjugate momentum).

When discussing orbit types in gauge theory, it is convenient to perform symplectic reduction in two stages. First, one factorizes with respect to the free action of pointed gauge transformations by applying regular symplectic reduction. Thereafter, one is left with an action of the compact gauge group $G$ on the quotient manifold which is not free (for a nonabelian $G$). That is, in the 
second stage one deals with singular symplectic reduction. Let us describe the first stage for the model under consideration. Given a lattice site $x_0$, it is not hard to see that the normal subgroup 
\beq
\label{G-ptgautrf}
\{g \in G^{\Lambda^0} : g(x_0) = \II\}\,,
\eeq
where $\II$ denotes the unit element of $G$, acts freely on $G^{\Lambda^1}$. Hence, one may pass to the quotient manifold and the residual action by the quotient Lie group of $G^{\Lambda^0}$ with respect to this normal subgroup. Clearly, the quotient Lie group is  naturally isomophic to $G$. The quotient manifold can be identified with a direct product of copies of $G$ and the quotient action can be identified with the action of $G$ by diagonal conjugation as follows. Choose a maximal tree $\mc T$ in the graph $\Lambda^1$ and define the tree gauge of $\mc T$ to be the subset
$$
\{a \in G^{\Lambda^1} : a(\ell) = \II ~\text{for all}~ \ell \in \mc T\}
$$
of $G^{\Lambda^1}$. One can readily see that every element of $G^{\Lambda^1}$ is conjugate under $G^{\Lambda^0}$ to an element in the tree gauge of $\mc T$ and that two elements in the tree gauge of $\mc T$ are conjugate under $G^{\Lambda^0}$ if they are conjugate under the action of $G$ via constant gauge transformations. This implies that the natural inclusion mapping of the tree gauge into $G^{\Lambda^1}$ descends to a $G$-equivariant diffeomorphism from that tree gauge onto the quotient manifold of $G^{\Lambda^1}$ with respect to the action of the subgroup \eqref{G-ptgautrf}. Finally, by choosing a numbering of the off-tree links in $\Lambda^1$, we can identify the tree gauge of $\mc T$ with the direct product of $N$ copies of $G$, where $N$ denotes the number of off-tree links. This number does not depend on the choice of $\mc T$. Then, the action of $G$ on the tree gauge via constant gauge transformations translates into the action of $G$ on $G^N$ by diagonal conjugation,
\beq\label{G-Wir-Q}
g \cdot \ul a = g \ul a g^{-1}\,,
\eeq
where $\ul a = (a_1 , \dots  , a_N) \in G^N$ and $g \in G$.
As a consequence, for the discussion of the role of orbit types we may pass in the first stage from the original Hamiltonian system with symmetries, given by the configuration space $G^{\Lambda^1}$, the symmetry group $G^{\Lambda^0}$ and the action \eqref{G-Wir-voll}, to the partially reduced Hamiltonian system with symmetries given by the configuration space 
$$
Q := G^N\,,
$$
the symmetry group $G$ and the action of $G$ on $Q$ given by diagonal conjugation \eqref{G-Wir-Q}. This is the system we will discuss here. As before, the classical phase space is given by the associated Hamiltonian $G$-manifold, whose underlying symplectic manifold is $\ctg Q$ endowed with its canonical cotangent bundle projection
$$
\pi: \ctg Q \to Q
\,.
$$
The action of $G$ on $Q$ naturally lifts to a symplectic action on $\ctg Q$ and the lifted action admits the standard momentum mapping
$$
\mm : \ctg Q \to \mf g^\ast
 \, , \quad
\mm(p)\big(X) := p(X_\ast)\,,
$$
where $p \in \ctg Q$, $X \in \mf g$ and $X_\ast$ denotes the Killing vector field defined by $X$. An easy calculation shows that under the global trivialization
\beq\label{G-trviz-N}
\ctg Q \equiv \ctg G^N \cong G^N \times \mf g^N
\eeq
induced by left-invariant vector fields and an invariant scalar product on $\mf g$, the lifted action is given by diagonal conjugation,
\beq
\label{G-Wir-P}
g \cdot (\ul a , \ul A)
 = 
\big(g a_1 g^{-1} , \dots , g a_N g^{-1} , \Ad(g) A_1 , \dots , \Ad(g) A_N\big)
\eeq
and the associated momentum mapping is given by 
\beq
\label{G-ImpAbb}
\mu(\ul a , \ul A)
 =
\sum_{i = 1}^N \Ad(a_i) A_i - A_i\,,
\eeq
see e.g.\ \cite[Section 10.7]{Buch}. The (fully) reduced phase space $\pha$ is obtained from $\ctg Q$ by singular symplectic reduction at $\mm = 0$, see Section \ref{Background}. That is, $\pha$ is the set of orbits of the lifted action of $G$ on the invariant subset $\mm^{-1}(0) \subseteq \ctg Q$, endowed with the quotient topology induced from the relative topology on this subset. In gauge theory, the condition $\mm=0$ corresponds to the Gau{\ss} law constraint. It can be shown that the action of $G$ on 
$\mm^{-1}(0)$ has the same orbit types as that on $Q$. As alread explained in Subsection \ref{Background}, the orbit type strata of $\pha$ are the connected components of the subsets of $\pha$ of elements with a fixed orbit type. By singular symplectic reduction, they are endowed with symplectic manifold structures. The bundle projection $\pi: \ctg Q \to Q$ induces a mapping $\pha \to Q/G$. This mapping is surjective, because $\mm$ is linear on the fibres of $\ctg Q$ and hence $\mm^{-1}(0)$ contains the zero section of $\ctg Q$. It need not preserve the orbit type though. For the study of various aspects of the above classical stratification we refer to \cite{cfg,cfgtop,FRS}, for a theory of 
singular (continuum) phase space reduction in the Fr\'echet context see \cite{DR} and further references therein.

In subsection \rref{A-OT} we will present the orbit types for $G=\SU(2)$.

%%%%%%%%%%%%%%%%%%%%%%%%%%%%%%%%%%%%%%%%%%%%%%%%%%%%%%%%%%%%%%%%%%%%%%%%%%%
%%%%%%%%%%%%%%%%%%%%%%%%%%%%%%%%%%%%%%%%%%%%%%%%%%%%%%%%%%%%%%%%%%%%%%%%%%%

\subsection{Canonical quantization. The observable algebra}
\label{CanQuant}

%%%%%%%%%%%%%%%%%%%%%%%%%%%%%%%%%%%%%%%%%%%%%%%%%%%%%%%%%%%%%%%%%%%%%%%%%%%
%%%%%%%%%%%%%%%%%%%%%%%%%%%%%%%%%%%%%%%%%%%%%%%%%%%%%%%%%%%%%%%%%%%%%%%%%%%%

The quantum theory is obtained via canonical quantization, see \cite{qcd2, qcd3, GR1, GR2}. As already explained in Subsection \ref{Q-picture}, the quantum reduction also goes in two stages. 
Using the fact that for the model at hand canonical quantization commutes with regular reduction \cite{qcd3}, we may limit our attention to the reduction after quantization 
of the model correponding to the singular reduction stage. 

By \eqref{G-trviz-N}, the classical partially reduced phase space will be identified with 
$\ctg G^N \cong G^N \times \mf g^N$ and the action of the group of local gauge transformations is reduced to the diagonal action of $G$ via \eqref{G-Wir-P}.   
To quantize the classical gauge connections, we generalize the Schr\"odinger representation for a particle on the real line acting on the Hilbert space $L^2(\RR)$ 
as follows: for any $\varphi \in L^2(G)$, we define the bounded operators
\beq
\label{G-GenSchrRep}
(U_g \varphi)(h) := \varphi(g^{-1}h) \, , \quad \big(T_f\varphi)(h):=  f(h)\varphi(h) \, ,
\eeq
where $g,\,h\in G$ and $f\in L^\infty(G)$. Here, $U$ is the left regular unitary representation of $G$ and $T$ is the natural representation of  $L^\infty(G) $ given by left multiplication. 
Clearly, $T$ and $U$ represent the position and momentum operator analogues, respectively. 
The pair $\pi_0 := (U,T)$ is referred to as the generalized Schr\"odinger representation. Below, it will be interpreted in the language of $C^\ast$-algebras. 
Note that $\pi_0$ is irreducible in the sense that the commutant
of ${U_G\cup T_{L^\infty(G)}}$ consists of the scalars. Also note that there is a natural ground state unit vector $\vp_0 \in L^2(G)$ which under the assumption that the Haar measure of $G$ is normalized is given by the constant
function $\vp_0 (h)=1$ for all $h\in G$.
Then,  $U_g \vp_0 = \vp_0$, and, by irreducibility, $\vp_0$ is cyclic with respect to the *-algebra generated by ${U_G\cup T_{L^\infty(G)}}$.
By construction, $\pi_0$ fulfils the intertwining relation
\beq
\label{G-IntertwR-UT}
U_g \circ T_f \circ  U^*_g = T_{\lambda_g(f)} \, ,
\eeq
where 
\beq
\label{G-leftact}
\lambda: G \to \Aut ( C(G))  \, , \quad
 \lambda_g(f)(h) := f(g^{-1} h) \, ,
\eeq
for any $g,h\in G$. This relation implies generalized commutation relations as follows.  
By \eqref{G-trviz-N}, the classical canonically conjugate momenta, also referred to as the colour electric fields, are given by elements of $\mf g$. For $X\in\mathfrak{g}$, we define the associated momentum operator by
\beq
\label{G-MomOp-L}
P_X: C^\infty(G) \to C^\infty(G) \, , \quad P_X\varphi: = i \ddtn U(\mr e^{tX})\varphi \, .
\eeq
Then, for any $f, \varphi \in C^\infty(G)$ and $X \in \mf g$,
$$
\big[P_X,\,T_f\big]\varphi = i \ddtn U(\mr e^{tX})T_f U(\mr e^{-tX}) \varphi   = i \ddtn T_{\lambda_{\exp(tX)}(f)} \varphi \, .
$$
Denoting the right invariant vector field on $G$ by $X^R$, we obtain
\beq
\label{G-GenCommRel}
\big[P_X,\,T_f\big] =  i T_{X^R(f)} \, .
\eeq
For $G = \RR$, this yields the standard Heisenberg commutation relations.
Since $P_X = \d U (X)$, we obtain a representation of the Lie algebra $\mathfrak{g}$ on $L^2(G)$ which obviously fulfils $P_X \vp_0=0$.

Now, the physical Hilbert space of the lattice quantum system is obtained by taking the tensor product of copies of $L^2(G)$ over all off-tree links, that is, 
\beq
\label{G-quHilbSp}
\rh := \bigotimes L^2(G) \cong L^2 (G^N) \, .
\eeq
Clearly, $\pi_0 = (U,T)$ induces a representation on $\rh$ denoted by $\pi := (\hat U, \hat  T)$. In detail,  for every off-tree link $\ell$, we define
\beq
\label{G-hat-T}
\hat{T}_f^{(\ell)}:= \otimes \cdots \otimes T^{(\ell)}_f \otimes \cdots \otimes \, ,
\eeq
and 
\beq
\label{G-hat-U}
\hat{U}_g^{(\ell)} := \otimes \cdots  \otimes {U}_g^{(\ell)} \otimes  \cdots \otimes \, ,
\eeq
where $T^{(\ell)}_f$ and $U^{(\ell)}_g$ are the multiplication and translation operators acting on the $\ell^{\rm th}$ tensor product factor of $\rh$, 
respectively. 

Next, recall that in the tree gauge the group of local gauge transformations simply coincides with $G$. We implement the gauge transformations law \eqref{G-Wir-P} at the quantum level as follows. We  take the unitary representation $V$ of $G$ on $L^2(G)$ given by 
$$
(V_g \varphi)(h) := \varphi(g^{-1}\,h\, g) \, , \quad \varphi \in L^2(G)\, ,
$$
and take the tensor product of this representation over all off-tree links to obtain a unitary representation on $\rh$, which we denote by the same letter.  
Clearly,  $V_g \vp_0 = \vp_0$ for all $g \in G$. Using  $V$, we define the local gauge transformations of the quantum fields from ${U_G \cup T_{L^\infty(G)}}$ by 
\beq
\label{G-GTr-Tf-1}
T_f \mapsto V_g \circ T_f \circ V_g^{-1} = T_{V_g f} \, ,  
\eeq
where $f\in L^\infty(G)\subseteq L^2(G)$, and 
\beq
\label{G-GTr-Tf-2}
U_h  \mapsto  V_g \circ U_h \circ V_g^{-1} =  U_{g \, h \, g^{-1}} \, ,
\eeq
for any $g\in G$. Moreover, since every operator $V_g$ preserves
the space $C^\infty(G)$, \eqref{G-GTr-Tf-2} implies 
\beq
\label{G-GTr-Tf-3}
V_g \circ P_X \circ V_g^{-1} = P_{\Ad (g) X } \, ,
\eeq
for any $ X\in\mf g$. 

Next, we construct the field algebra and the observable algebra of our model. 
For functional analytic basics used below we refer to \cite{Pedersen} or \cite{Blackadar}.
By \eqref{G-IntertwR-UT}, 
the generalized Schr\"odinger representation $\pi_0 = (U,T) $ is a covariant representation of the $C^\ast$-dynamical system $(C(G), G, \lambda)$ with $\lambda: G \to \Aut \big(C(G)\big)$ defined by \eqref{G-leftact}. Associated with this $C^\ast$-dynamical system, there is a natural crossed product $C^\ast$-algebra $C(G) \rtimes_\lambda G $.
Its representations are exactly the
covariant representations of the $C^*$-dynamical system defined by $\lambda$.
It is well known that $C(G) \rtimes_\lambda G$ is isomorphic to the algebra of compact operators on $L^2(G)$, 
\beq
\label{G-Iso-CrProdAlg}
C(G) \rtimes_\lambda G \cong {\mathfrak K} \big( L^2(G) \big) \, , 
\eeq 
see Theorem II.10.4.3 in \cite{Blackadar}.
In fact, 
$$
\pi_0 \big(C(G) \rtimes_\lambda G\big) = {\mathfrak K} \big(L^2(G)\big) \, .
$$
Since ${\mathfrak K}\big(L^2(G)\big)$ has a unique irreducible representation up to unitary equivalence, it follows that $\pi_0$ is the unique irreducible
covariant representation of $(C(G),G,\lambda)$ (up to equivalence). Moreover, as $\vp_0$ is cyclic for
${\mathfrak K} \big(L^2(G)\big)$, $\pi_0$ is unitarily equivalent to the GNS-representation
of the vector state $\omega_0$ given by $\omega_0 (A):={(\vp_0,\pi_0(A)\vp_0)}$ for $A \in C(G)\rtimes_\lambda G$.
By taking the tensor product of copies of $ C(G) \rtimes_\lambda G$ over all off-tree links, we obtain the following field algebra:
\beq
\label{G-FieldAlg}
\mf A_0
 := 
\bigotimes\big(C(G) \rtimes_\lambda G \big) 
 \cong  
\bigotimes {\mathfrak K} ( L^2(G))
\,.
\eeq
Since $\mf A_0$ is simple, there is only one irreducible representation, up to unitary equivalence. Moreover, all representations are faithful. This implies the following.

\bsz
\label{S-Rep-FA}

The field algebra $\mf A_0$ is faithfully and irreducibly represented by 
$\big(\rh, \pi\big) $, that is, 
\beq
\label{G-Rep-FA}
\pi (\mf A_0) = {\mathfrak K} \big(\rh\big) \, .
\eeq

\esz

Finally, we note that we have a natural action $\alpha$ of the (reduced) group of local gauge transformations $G$ on $\mf A_0$ by automorphisms. In the representation $\pi$ it is given by $g \to {\rm Ad}(V_g)$ via \eqref{G-GTr-Tf-1} and \eqref{G-GTr-Tf-2}. This action clearly preserves $\pi \big(\mf A_0\big) = \mf K(\rh)$ and, since  
$g  \to V_g$ is strongly operator continuous, it defines a strongly continuous action $\alpha$ of $G$ on $\pi \big(\mf A_0\big)$ and, thus, on $\mf A_0$. For a detailed discussion of this action we refer to \cite{GR2}, \cite{RS-Buch2}.

Now, we can define the observable algebra of the system; see \cite{GR2}, \cite{RS-Buch2} for the details.

\bdf[Observable algebra]\label{D-ObsAlg}

The observable algebra of the lattice gauge theory is defined by 
$$
\mf A := \mf A_0^G /(\mf I \cap \mf A_0^G)
\,,
$$
where $\mf A_0^G \subseteq \mf A_0$ is the subalgebra of gauge invariant elements and  $\mf I  \subseteq \mf A_0 $ is the ideal generated by the local Gau{\ss} laws.

\edf

Recall that, under the representation $\pi$, the field algebra $\mf A_0$ gets identified with ${\mathfrak K} (\rh)$. Under this identification, the subalgebra $\mf A_0^G$ can be viewed as the commutant of the unitary representation $V$ in $\mf K(\rh)$. Consider the closed subspace $\rrh := \rh^G \subseteq \rh$ consisting of $G$-invariant vectors, 
\beq
\label{G-Hinv}
\rrh
 := 
\{ \Phi \in \rh \, \, | \, \, V_g ( \Phi) = \Phi \, \, \, \text{for all} \,  \, \, g \in G \} \ .
\eeq

\btm
\label{T-ObsAlg-Rep}

The  observable algebra $\mf A$ is isomorphic to the algebra of compact operators on $\rrh$, 
\beq\label{G-ObsAlg-Rep}
\mf A \cong \mf K \big(\rrh\big)
\,.
\eeq

\etm

For the proof, see \cite{qcd3}. Observe that by passing from the field algebra $\mf A_0$ to the observable algebra $\mf A$ we reduce the remaining gauge symmetry implemented 
by the diagonal action of $G$ at the quantum level. Finally, we recall that in Example 
\ref{GaugeTh} we have outlined that $\mf A$ may be obtained via the $T$-procedure.

%%%%%%%%%%%%%%%%%%%%%%%%%%%%%%%%%%%%%%%%%%%%%%%%%%%%%%%%%%%%%%%%%%%%%%%%%%%%%
%%%%%%%%%%%%%%%%%%%%%%%%%%%%%%%%%%%%%%%%%%%%%%%%%%%%%%%%%%%%%%%%%%%%%%%%%%%

\subsection{K\"ahler quantization and reduction}
\label{Quantization-Reduction}

%%%%%%%%%%%%%%%%%%%%%%%%%%%%%%%%%%%%%%%%%%%%%%%%%%%%%%%%%%%%%%%%%%%%%%%%%%%
%%%%%%%%%%%%%%%%%%%%%%%%%%%%%%%%%%%%%%%%%%%%%%%%%%%%%%%%%%%%%%%%%%%%%%%

In the next step, we wish to implement the classical gauge orbit stratification at the quantum level. Recall the general procedure from Subsection \ref{AlgObs-CostrHSp}. Here, the construction will be accomplished by using the (generalized) polar mapping. For that purpose, let $\mf g_\CC$ denote the complexification of $\mf g$ and let $G_\CC$ denote the complexification of $G$. This is a complex Lie group having $G$ as its maximal compact subgroup. It is unique up to isomorphisms. For $G = \SU(n)$, we have $G_\CC = \SL(n,\CC)$. By restriction, the exponential mapping  
$$
\exp : \mf g_\CC \to G_\CC
$$
of $G_\CC$ and multiplication in $G_\CC$ induce a diffeomorphism
\beq\label{G-poldec-1}
G\times\mf g \to G_\CC
 \,, \quad 
(a,A) \mapsto a\exp(\mr i A)\, ,
\eeq
which is equivariant with respect to the action of $G$ on $G\times \mf g$ by 
$$
g \cdot (a,A) := \big(g a g^{-1} , \Ad(g) A\big)
$$
and the action of $G$ on $G_\CC$ by conjugation. For $G = \SU(n)$, this diffeomorphism amounts to the inverse of the polar mapping. By composing this diffeomorphism with the left global cotangent bundle trivialization, we obtain a diffeomorphism 
\beq\label{G-dfm-P}
\ctg G^N \to G_\CC^N \, ,
\eeq 
which is equivariant with respect to the lifted action of $G$ on $\ctg G^N$ and the action of $G$ on $G_\CC^N$ by diagonal conjugation. Via this diffeomorphism, the complex structure of $G_\CC^N$ and the symplectic structure of $\ctg G^N$ combine to a K\"ahler structure. Now, half-form K\"ahler quantization  on $G_\CC^N$ yields the Hilbert space 
$$
{\cal K} := HL^2(G_\CC^N , \mr d \nu_\hbar)
$$
of holomorphic functions on $G_\CC^N$ which are square-integrable with respect to the measure 
\beq
\label{measure-nu}
\mr d \nu_\hbar = \mr e^{- \kappa / \hbar} \, \eta \, \ve\,,
\eeq
where $\kappa$ is the K\"ahler potential on $G_\CC^N$, $\eta$ is the half-form correction and $\ve$ is the Liouville measure on $\ctg G^N$. See \cite{Hall:cptype} for details. Reduction then yields the closed subspace 
\beq
\label{hol-Rep-inv}
\rkh := {\cal K}^G \equiv HL^2(G_\CC^N , \mr d \nu_\hbar)^G
\eeq
of $G$-invariants as the Hilbert space of the reduced system.

\bbm\label{Segal-Bargmann}

The above result belongs to Hall \cite{Hall:cptype}. Alternatively, as also observed by Hall, the Hilbert space $HL^2(G_\CC^N , \mr d \nu)$ is obtained via the Segal-Bargmann transformation for compact Lie groups \cite{Hall:SBT}. In more detail, the Segal-Bargmann transformation 
$$
\Phi: L^2 (G^N) \to  HL^2(G_\CC^N , \mr d \nu_\hbar)
$$
is a unitary isomorphism, which restricts to a unitary isomorphism of the subspaces of invariants. That is, under this isomorphism, the canonical representation space 
$\rrh \equiv L^2(G^N)^G$ of the observable algebra $\mathfrak A$ gets identified with the Hilbert space $\rkh \equiv HL^2(G_\CC^N , \mr d \nu_\hbar)^G$.  
\qeb

\ebm

%%%%%%%%%%%%%%%%%%%%%%%%%%%%%%%%%%%%%%%%%%%%%%%%%%%%%%%%%%%%%%%%%%%%%%%%%%%

\subsection{Orbit type costratification}

%%%%%%%%%%%%%%%%%%%%%%%%%%%%%%%%%%%%%%%%%%%%%%%%%%%%%%%%%%%%%%%%%%%%%%%%%%%

According to Remark \ref{Segal-Bargmann}, the Hilbert space $\rkh$ given by \eqref{hol-Rep-inv} may be identified with the canonical representation space of the observable algebra $\mf A$. Here, we adopt this point of view. By the construction presented in Subsection 
\ref{AlgObs-CostrHSp}, the subspaces associated with the orbit type strata of $\pha$ are, by definition, the orthogonal complements of the subspaces of functions vanishing on those strata. We first have to clarify how to interpret elements of $\rkh$ as functions on $\pha$. In the case $N=1$ discussed in \cite{HRS}, this is readily done by observing that $\pha \cong T_\CC/W$, where $T$ is a maximal torus in $G$ and $W$ the corresponding Weyl group, and by using the isomorphism 
$$
HL^2(G_\CC , \mr d \nu)^G \cong HL^2(T_\CC , \mr d \nu_T)^W \, .
$$ 
Here, the measure $\mr d \nu_T$ is obtained from $\mr d \nu$ by integration over the conjugation orbits in $G_\CC$, thus yielding an analogue of Weyl's integration formula for $HL^2(G_\CC , \mr d \nu)$. In the general case, the argument is as follows. 

Consider the action of $G_\CC$ on $G_\CC^N$ by diagonal conjugation. For $\ul a \in G_\CC^N$, let $G_\CC \cdot \ul a$ denote the corresponding orbit. Since $G_\CC$ is not compact, $G_\CC \cdot \ul a$ need not be closed. If a holomorphic function on $G_\CC^N$ is invariant under the action of $G$ by diagonal conjugation, then it is also invariant under the action of $G_\CC$ by diagonal conjugation, i.e., it is constant on the orbit $G_\CC \cdot \ul a$ for every $\ul a \in G_\CC^N$. Being continuous, it is then constant on the closure $\ol{G_\CC \cdot \ul a}$. As a consequence, it takes the same value on two orbits whenever their closures intersect. We may therefore pass to orbit closure equivalence classes: two elements $\ul a , \ul b \in G_\CC^N$ are said to be orbit closure equivalent if there exist $\ul c_1 , \dots , \ul c_r \in G_\CC^N$ such that
$$
\ol{G_\CC \cdot \ul a} \cap \ol{G_\CC \cdot \ul c_1} \neq \varnothing
 \,,~~
\ol{G_\CC \cdot \ul c_1} \cap \ol{G_\CC \cdot \ul c_2} \neq \varnothing
 \,,\, \dots \,,~~
\ol{G_\CC \cdot \ul c}_r \cap \ol{G_\CC \cdot \ul b} \neq \varnothing\,.
$$
Let $G_\CC^N /\!/ G_\CC$ denote the topological quotient. This notation is motivated by the fact that the quotient provides a categorical quotient of $G_\CC^N$ by $G_\CC$ in the sense of geometric invariant theory \cite{Mumford}. By construction, the elements of $\rkh$ descend to continuous functions on $G_\CC^N /\!/ G_\CC$.

\btm
\label{T-KN}

The natural inclusion mapping $\mu^{-1}(0) \to G_\CC^N$ induces a homeomorphism 
\beq\label{G-KN-Hoem}
\pha \to G_\CC^N /\!/ G_\CC\,.
\eeq

\etm

For the general arguments, see \cite{HeinznerLoose}. For details in the context of our model, see \cite{FuRS}.

As a result, via the homeomorphism \eqref{G-KN-Hoem}, the elements of $\rkh$ can be interpreted as functions on $\pha$. Now, according to the construction presented in Section \rref{AlgObs-CostrHSp}, to a given orbit type stratum $\pha_\tau \subseteq \pha$, there corresponds the closed subspace 
$$
\mc V_{\tau} := \{\psi \in \rkh : \psi_{\res \pha_\tau} = 0\}
$$
and the subspace $\rkh_\tau$ associated with $\pha_\tau$ is given by the orthogonal complement of $\mc V_\tau$ in $\rkh$. Thus, we have the orthogonal decomposition
$$
\rkh = \rkh_\tau \oplus \mc V_\tau\,.
$$

In the remainder we apply the general theory of Section \ref{Str-QS} to the model discussed 
here for the case $G = \SU(2)$.

%%%%%%%%%%%%%%%%%%%%%%%%%%%%%%%%%%%%%%%%%%%%%%%%%%%%%%%%%%%%%%%%%%%%%%%%%%%
%%%%%%%%%%%%%%%%%%%%%%%%%%%%%%%%%%%%%%%%%%%%%%%%%%%%%%%%%%%%%%%%%%%%%%%%%%%%%

\subsection{Orbit type strata for $G=\SU(2)$}
\label{A-OT}

%%%%%%%%%%%%%%%%%%%%%%%%%%%%%%%%%%%%%%%%%%%%%%%%%%%%%%%%%%%%%%%%%%%%%%%%%%%
%%%%%%%%%%%%%%%%%%%%%%%%%%%%%%%%%%%%%%%%%%%%%%%%%%%%%%%%%%%%%%%%%%%%%%%%%%%%%

\label{B-OT-N}

We start with deriving the orbit type strata for $G = \SU(2)$. For details, see \cite{FuRS}. Let $Z$ denote the center of $G$ and let $T \subseteq G$ denote the subgroup of diagonal matrices. Clearly, $T$ is a maximal toral subgroup, isomorphic to $\mr U(1)$. Let $\mf t \subseteq \mf g$ be the Lie subalgebra associated with $T$. 

First, the case $N=1$ has been discussed already in Subsection \ref{Background}. 
Thus, let us consider the case $N>1$. The stabilizer of an element $(\ul a , \ul A) \in G^N \times \mf g^N$ is given by 
$$
\mr C_G(a_1) \cap \cdots \cap \mr C_G(a_N) \cap \mr C_G(A_1) \cap \cdots \cap \mr C_G(A_N)\,,
$$
where $\mr C_G(\cdot)$ denotes the respective centralizer in $G$, i.e., 
$$
\mr C_G(a_i) = \{g \in G : g a_i g^{-1} = a_i\}
 \,,~~~~~~
\mr C_G(A_i) = \{g \in G : \Ad(g) A_i = A_i\}\,.
$$
The centralizer $\mr C_G(a_i)$ is conjugate to $T$ unless $a_i = \pm\II$, where $\mr C_G(\pm\II) = G$. Similarly, the centralizer $\mr C_G(A_i)$ is conjugate to $T$ unless $A_i = 0$, where $\mr C_G(0) = G$. Since two distinct subgroups which both are conjugate to $T$ intersect in the center $Z$, by taking intersections, we see that the stabilizer can be $G$, conjugate to $T$, or $Z$, where $Z$ is the generic situation. Accordingly, there are three orbit types and these can be labeled by $G$, $T$ and $Z$, where $Z$ is the principal orbit type. The corresponding orbit type subsets of $G^N \times \mf g^N$ are as follows:

\ben

\item[$(G)$] Here, $\mr C_G(a_i) = \mr C_G(A_i) = G$ for all $i$. Hence, $(\ul a , \ul A)$ has orbit type $G$ iff 
$$
(\ul a , \ul A) \in Z^N \times \{0\}^N\,.
$$

\item[$(T)$] Up to conjugacy, one of the centralizers $\mr C_G(a_i)$ or $\mr
C_G(A_i)$ equals $T$ and all the other centralizers contain $T$. If $\mr C_G(a_j)$ contains $T$, then $a_j \in T$. If $\mr C_G(A_j)$ contains $T$, then $A_j \in \mf t$. Hence, $(\ul a , \ul A)$ has orbit type $T$ iff it is conjugate to an element of the subset 
$$
\left(T^N \times \mf t^N\right) \setminus \left(Z^N \times \{0\}^N\right)\,.
$$

\item[$(Z)$] The point $(\ul a , \ul A)$ has orbit type $Z$ iff it does not have
obit type $T$ or $G$, i.e., iff it is not conjugate to an element of $T^N \times
\mf t^N$.  

\een

To find the corresponding orbit type subsets of $\pha$, we have to intersect the orbit type subsets with the momentum level set $\mu^{-1}(0)$ and pass to $G$-orbits. According to \eqref{G-ImpAbb}, 
$$
T^N \times \mf t^N \subset \mu^{-1}(0)\,.
$$
Since $\mu^{-1}(0)$ is $G$-invariant, this implies that the subsets of orbit type $T$ and orbit type $G$ are contained in $\mu^{-1}(0)$. Since they do not exhaust $\mu^{-1}(0)$, all three orbit types survive the reduction procedure, thus yielding three orbit type subsets of $\pha$. Finally, to find the orbit type strata, we have to decompose these orbit type subsets into connected components:

\ben

\item[$(G)$] Since the elements of $Z^N \times \{0\}^N$ are invariant under the action of $G$, each of them projects to a single point in $\pha$. Therefore, there exist $2^N$ orbit type strata of orbit type $G$, each of which consists of a single point representing the (trivial) orbit  of an element of $Z^N \times \{0\}^N$. Since such an element is of the form $(\nu_1 \II , \dots , \nu_N \II , 0, \dots , 0)$ for some sequence of signs $\ul\nu = (\nu_1 , \dots , \nu_N)$, we denote the corresponding stratum by $\pha_{\ul\nu}$.  

\item[$(T)$] Since $Z^N \times \{0\}^N$ consists of finitely many points and
$T^N \times \mf t^N$ has dimension at least $2$, the complement $(T^N \times \mf
t^N) \setminus (Z^N \times \{0\}^N)$ is connected. Since the subset of $\pha$ of
orbit type $T$ is the image of 
$$
(T^N \times \mf t^N) \setminus (Z^N \times \{0\}^N)
$$
under the natural projection $\mu^{-1}(0) \to \pha$, it is connected,
too. Hence, it forms an orbit type stratum, $\pha_T$.

\item[$(Z)$] Since $\mf g^\ast$ has dimension $3$, the level set $\mu^{-1}(0)$ generically has dimension $2N \cdot 3 - 3 = 3(2N-1)$. On the other hand, since $T$ has dimension $1$ and the elements of $T^N \times \mf t^N$ have stabilizer $T$ under the action of $G$, the subset of $G^N \times \mf g^N$ of orbit type $T$ has dimension $2N \cdot 1 + (3-1) = 2(N+1)$. Hence, if the orbit type $Z$ occurs in $\pha$, i.e., if $N \geq 2$, then the subset of $\mu^{-1}(0)$ generated from $T^N \times \mf t^N$ by the action of $G$ has codimension
$$
3(2N-1) - 2(N+1) = 4N-5 \geq 3\,.
$$
Therefore, its complement is connected. Since the complement coincides with the subset of $\mu^{-1}(0)$ of orbit type $Z$, the subset of $\pha$ of this orbit type is connected. Hence, it forms an orbit type stratum, $\pha_Z$.
\qeb

\een

%%%%%%%%%%%%%%%%%%%%%%%%%%%%%%%%%%%%%%%%%%%%%%%%%%%%%%%%%%%%%%%%%%%%%%%%%%%
%%%%%%%%%%%%%%%%%%%%%%%%%%%%%%%%%%%%%%%%%%%%%%%%%%%%%%%%%%%%%%%%%%%%%%%%%%%%%

\subsection{Costratified Hilbert space and stratified observable algebra for $G=\SU(2)$}
\label{Strat-obsAlg}

%%%%%%%%%%%%%%%%%%%%%%%%%%%%%%%%%%%%%%%%%%%%%%%%%%%%%%%%%%%%%%%%%%%%%%%%%%%
%%%%%%%%%%%%%%%%%%%%%%%%%%%%%%%%%%%%%%%%%%%%%%%%%%%%%%%%%%%%%%%%%%%%%%%%%%%%%

In this section, we construct the closed subspaces $\rkh_\tau \subset \rkh$ and the $C^\ast$-subalgebras $\mf A_\tau \subset \mf A$ associated with the classical phase space strata $\pha_\tau$ for $G=\SU(2)$. As discussed in subsection \rref{A-OT}, here $\tau=\ul\nu, T$. According to Sections \rref{CanQuant} and \rref{Quantization-Reduction}, the Hilbert space is given by $\rkh = HL^2(G_\CC^N,\mr d \nu_\hbar)^G$ and the observable algebra is given by $\mf A = \mf K(\rkh)$. The invariant representative functions form a dense subspace of $\rkh$ which is closed under pointwise multiplication of functions and which satisfies the denseness condition for every $\tau$, see Proposition 3.4 in \cite{FuRS}. Hence, we may take them as the dense algebra $\mc R$ needed in the construction presented in Section \rref{str-AlgObs}.

Since the elements of $\rkh$ are given by functions on $G_\CC^N$, vanishing of $\xi \in \rkh$ on the stratum $\pha_\tau$ in fact means that $\xi_{\res (G_\CC^N)_\tau} = 0$, where $(G_\CC^N)_\tau \subset G_\CC^N$ is the subset which under the natural projection $G_\CC^N \to G_\CC^N // G_\CC$ and the homeomorphism \eqref{G-KN-Hoem} corresponds to the orbit type stratum $\pha_\tau \subset \pha$. Thus, relations defining $\pha_\tau$ are given by relations defining the subsets $(G_\CC^N)_\tau$. Since the subspaces $\rkh_\tau$ and the subalgebras $\mf A_\tau$ can most conveniently be described in terms of the associated orthogonal projections $\mr p_\tau : \rkh \to \rkh_\tau$, it is these projections which have to be found. 
\bigskip

First, consider the case $\tau = T$.

\btm\label{T-SU2-defrel-T}~

The $G$-invariant representative functions
 \ala{
r^T_{ij}(\ul a) & := \tr\big([a_i,a_j]^2\big) \,,\qquad 1 \leq i < j \leq N 
\phantom{< k\,.}
\\
r^T_{ijk}(\ul a) & := \tr\big([a_i,a_j]a_k\big) \,,\qquad 1 \leq i < j < k \leq N
 }
satisfy the zero locus condition wrt.\ the topological closure of $(G_\CC^N)_T$ and the radical ideal condition.

\etm

\bpf

See \cite{FuRS}, Theorems 5.2 and 6.1.
\epf

The associated projection $\mr p_T : \rkh \to \rkh_T$ can be obtained by a perturbative analysis of the relations given in the theorem. This is work in progress and will be discussed elsewhere. As a first step, in \cite{FJRS} we have determined the expansion coefficients of the products $r^T_{ij} \psi_\alpha$ and $r^T_{ijk} \psi_\alpha$ for $\{\psi_\alpha\}$ being the orthonormal basis of quasi-characters.
\bigskip

Now, consider the case $\tau = \ul\nu$.

\btm\label{T-SU2-defrel-Z}

For every sequence of signs $\ul\nu$, the $G$-invariant representative functions given in Theorem \rref{T-SU2-defrel-T}, together with 
 \ala{
r^{\ul\nu}_i(\ul a) & := \tr(a_i) - 2 \nu_i 
\,,\qquad 
& & 1 \leq i \leq N 
\,,
\\
r^{\ul\nu}_{ij}(\ul a) & := \tr(a_i a_j) - 2 \nu_i \nu_j
\,,\qquad 
& & 1 \leq i < j \leq N
\,,
 }
satisfy the zero locus condition wrt.\ $(G_\CC^N)_{\ul\nu}$ and the radical ideal condition.

\etm

\bpf

Let $\ul\nu$ be given. In \cite{FuRS}, Theorem 5.3, it was shown that the zero locus condition holds without the functions $r^{\ul\nu}_{ij}$. In the proof of that theorem it has furthermore been shown that the conditions $r^T_{ij}(\ul a) = 0$, $r^T_{ijk}(\ul a) = 0$ and $r^{\ul\nu}_i(\ul a) = 0$ imply that 
$$
a_i = \bbma \nu_i & \beta_i \\ 0 & \nu_i \ebma 
$$
with some $\beta_i \in \CC$ for all $i$. Hence, the functions $r^{\ul\nu}_{ij}$ vanish on $\pha_{\ul\nu}$, so that adding them does not destroy the zero locus condition. 

To check the radical ideal condition, it suffices to show that the ideal $\mc I$ generated in $\mc R$ by the functions given in the theorem coincides with the vanishing ideal $\mc N$ of the point $(\nu_1 \II , \dots , \nu_N \II)$. By the zero locus condition, $\mc I \subset \mc N$. To prove the converse inclusion, let $\mc I_0$ denote the ideal in $\mc R$ generated by the $r^T_{ij}$ and the $r^T_{ijk}$. For $1 \leq i,j \leq N$, define $G$-invariant representative functions $t_i$, $t_{ij}$ by
$$
t_i(\ul a) := \tr(a_i)
\,,\qquad
t_{ij}(\ul a) := \tr(a_i a_j)
\,.
$$
It has been shown in the proof of Lemma 6.3 in \cite{FuRS} that the functions 
\beq\label{G-T-SU2-defrel-Z-1}
t_{i_1} \cdots t_{i_r} t_{k_{11},k_{12}} \cdots t_{k_{s1},k_{s2}}
\eeq
with $i_1 \leq \cdots \leq i_r$ and $(k_{11},k_{12}) < \cdots < (k_{s1},k_{s2})$ (lexicographic ordering), and where $r,s=0,1,2,\dots$, form a basis in a vector space complement of $\mc I_0$ in $\mc R$. Here, in case $r=0$ or $s=0$, the corresponding factor collapses to $1$. Resolving
$$
t_i = r^{\ul\nu}_i + 2 \nu_i
\,,\qquad
t_{ij} = r^{\ul\nu}_{ij} + 2 \nu_i \nu_j
$$
and plugging this into \eqref{G-T-SU2-defrel-Z-1}, we see that every such basis element can be written as a linear combination of the constant function $1$ and an element of $\mc I$. It follows that $1$ spans a vector space complement of $\mc I$ in $\mc R$ and hence that $\mc N \subset \mc I$.
\epf

Since each of the strata $\pha_{\ul\nu}$ consists of a single point, the closed subspaces $\rkh_{\ul\nu}$, and thus the associated projections $\mr p_{\ul\nu} : \rkh \to \rkh_{\ul\nu}$, can be determined explicitly without using the relations by the following direct argument. 

Let $\{\psi_\alpha : \alpha \in A\}$ be an orthonormal basis of $\rkh$ consisting of invariant representative function which contains a constant function $\psi_0$. Such a basis exists, because the constant functions are invariant representative functions and they belong to $HL^2(G_\CC^N,\mr d\nu)$. For example, one may use the orthonormal basis of quasi-characters presented in \cite{FJRS}.

\btm\label{T-SU2-OP}

For every sequence $\ul\nu$ of signs, the subspace $\rkh_{\ul \nu}$ is spanned by the single element
$$
\psi_{\ul \nu} 
 =
\sum_{\beta \in A}
 ~
\ol{\psi_\beta(\nu_1 \II , \dots , \nu_N \II)} \, \psi_\beta
\,.
$$

\etm

\bpf

Since for an invariant function $\psi$, the condition to vanish on $(G_\CC^N)_{\ul\nu}$ is equivalent to the condition $\psi(\nu_1 \II , \dots , \nu_N \II) = 0$, the vanishing subspace $\mc V_{\ul \nu}$ is spanned by
\beq\label{G-SU2-OP-1}
\vp_\alpha := \psi_\alpha - \psi_\alpha(\nu_1 \II , \dots , \nu_N \II) \, 1
 \,,~~~~~~
\alpha \in A\,, ~ \alpha \neq 0\,,
\eeq
where $1$ denotes the constant function with value $1$. Since this function complements the functions \eqref{G-SU2-OP-1} to a basis in $\rkh$, the subspace $\rkh_{\ul\nu}$ has dimension $1$. Hence, it suffices to show that $\psi_{\ul\nu}$ is orthogonal to the functions \eqref{G-SU2-OP-1}. Denoting the scalar product in $\rkh$ by $\langle \cdot , \cdot \rangle$ and writing $\ul b = (\nu_1 \II , \dots , \nu_N \II)$, for given $\alpha \in A$ we compute
 \ala{
\langle \psi_{\ul \nu} , \vp_\alpha \rangle
 = &
\sum_{\beta \in A}
\psi_\beta(\ul b)
\langle \psi_\beta , \psi_\alpha \rangle
 -
\sum_{\beta \in A}
\psi_\beta(\ul b)
\psi_\alpha(\ul b)
\langle \psi_\beta , 1 \rangle
\,.
 }
Since the basis is orthonormal, the first sum yields $\psi_\alpha(\ul b)$. Moreover, since $\psi_0$ is constant, $\langle \psi_\beta , 1 \rangle = 0$ unless $\beta = 0$. Hence, the second sum reduces to 
$$
\psi_0(\ul b)
\psi_\alpha(\ul b)
\langle \psi_0 , 1 \rangle
 =
\psi_\alpha(\ul b)
\langle \psi_0 , \psi_0(\ul b) 1 \rangle\,.
$$
Since $\psi_0(\ul b) \, 1 = \psi_0$, the scalar product gives $1$. Hence, 
$
\langle \psi_{\ul \nu} , \vp_\alpha \rangle = 0
$
for all $\alpha \in A$, indeed.
\epf

\bfg\label{F-SU2-OP}

For every sequence $\ul\nu$ of signs, the projection associated with the stratum $\pha_{\ul\nu}$ is given by 
\eqqedan
\beq
\mr p_{\ul\nu}
 = 
\frac{\ket{\psi_{\ul\nu}} \bra{\psi_{\ul\nu}}}{\|\psi_{\ul\nu}\|^2}
\,.
\eqqed
\eeq
\eqqedaus
\efg

\bfg\label{F-SU2-OA}

For every sequence $\ul\nu$ of signs, the observable algebra associated with the stratum $\pha_{\ul\nu}$ is given by $\mf A_{\ul\nu} = \CC\mr p_{\ul\nu}$.

\efg

\bpf

Since $\mr p_{\ul\nu}$ has finite rank, it belongs to $\mf A \equiv \mf K(\rkh)$. Hence, 
$$
\mf A_{\ul\nu} = \mr p_{\ul\nu} \mf A \mr p_{\ul\nu}
\,,
$$
and every element of $\mf A_{\ul\nu}$ is of the form
$$
\mr p_{\ul\nu} a \mr p_{\ul\nu}
 =
 \frac{
\langle \psi_{\ul\nu} | a \psi_{\ul\nu} \rangle \psi_{\ul\nu}
 }{
\|\psi_{\ul\nu}\|^4
 }
 =
\langle a \rangle_{\psi_{\ul\nu}} \, \mr p_{\ul\nu}
\,,
$$
where $\langle a \rangle_{\psi_{\ul\nu}}$ denotes the expectation value of the observable $a$ in the state $\psi_{\ul\nu}$.
\epf

\bbm
\label{Bem-SU2-D}

The complementary projection $\mr q_{\ul\nu} = \II_{\rkh} - \mr p_{\ul\nu}$ defining the hereditary $C^\ast$-subalgebra $\mf D_{\ul\nu} = (\mr q_\tau |\mf A|_s \mr q_\tau) \cap \mf A$ has infinite rank and thus does not belong to $\mf A \equiv \mf K(\rkh)$. 
\qeb

\ebm

\bbm
\label{Bem-SU2}

Here, we take up the case $N=1$, which was already discussed in Section \rref{Background}. According to subsection \rref{A-OT}, there are two secondary strata $\pha_\pm$ which under the homeomorphism with $G_\CC//G_\CC$ correspond to (the orbit closure equivalence classes of) the isolated points $\pm \II$. Hence, we may apply Theorem \rref{T-SU2-OP}. As an orthonormal basis, we may choose the normalized characters of $G_\CC$,
$$
(\hbar\pi)^{-\frac{3}{4}}e^{-\hbar\beta(n+1)^2/2} \, \chi^\CC_n
\,,\qquad
n=0,1,2,\dots
\,.
$$
Here, $n$ is twice the spin and $\beta$ is a scaling parameter for the invariant scalar product on $\su(2)$. Clearly, $\chi^\CC_n(\pm \II) = (\pm 1)^n (n+1)$. Hence, $\rkh_\pm$ is spanned by the vector
$$
\psi_\pm
 =
\sum_{n=0}^\infty (\pm 1)^n \, (n+1) \, \mr e^{-\hbar \beta^2 (n+1)^2}\chi^\CC_n
\,.
$$
One can check that 
$$
\|\psi_\pm\|^2 = \sum\limits^\infty_{n=1} n^2 e^{-\hbar \beta^2 n^2}
\,.
$$
Thus, according to Corollary \rref{F-SU2-OP}, the projection associated with the stratum $\pha_\pm$ is given by 
$$
\mr p_\pm
 = 
\sum\limits_{n,m=0}^\infty 
 \frac{
(\pm 1)^{n+m} \, (n+1)(m+1) \, \mr e^{-\hbar \beta^2 ((n+1)^2 + (m+1)^2)} 
 }{
\sum\limits^\infty_{n=1} n^2 e^{-\hbar \beta^2 n^2}
 }
 ~
\ket{\chi^\CC_n} \bra{\chi^\CC_m}
\,\,.
$$
Since $\pha_\pm$ are the only secondary strata, according to Corollary \rref{F-SU2-OA}, the stratification of the observable algebra $\mf A$ consists of the two one-dimensional subalgebras $\mf A_+ = \CC \mr p_+$ and $\mf A_- = \CC \mr p_-$.
\qeb

\ebm

\bbm

For $\alpha \in A$, $\alpha \neq 0$, let $\vp_\alpha$ denote the function defined by \eqref{G-SU2-OP-1} and put
$$
\tilde\vp_\alpha := \frac{\vp_\alpha}{\|\vp_\alpha\|}
 \,,\qquad
\mr q_\alpha := \ket{\tilde\vp_\alpha} \bra{\tilde\vp_\alpha}
\,.
$$
We show that 
$$
\mf C := \{\mr q_\alpha : \alpha \in A , \alpha \neq 0\}
$$
is a self-adjoint generating set of annihilating operators for $\pha_{\ul\nu}$ and can thus be taken as the constraint set for the $T$-procedure. Self-adjointness is obvious and the annihilation property holds by construction. To check the generation property, let $\mr q$ be a one-dimensional projection onto a subspace spanned by an element $\xi$ of $\mc R$ vanishing on $\pha_\tau$. Assuming $\xi$ to be normalized, we have $\mr q = \ket\xi \bra\xi$. Expanding $\xi$ wrt.\ the basis made up by $\psi_0$ and the $\vp_\alpha$, $\alpha \neq 0$, we see that 
$$
\xi = \sum_{\alpha \neq 0} C_\alpha \, \vp_\alpha
$$
for certain $C_\alpha \in \CC$. Therefore, 
$$
\mr q
 = 
\sum_{\alpha,\alpha' \neq 0} C_\alpha \ol{C_{\alpha'}}
 ~
\ket{\vp_\alpha} \bra{\vp_{\alpha'}}
 =
\sum_{\alpha,\alpha' \neq 0} C_\alpha \ol{C_{\alpha'}} 
 ~
\mr q_{\alpha'}
 \Big(
\ket{\vp_\alpha} \bra{\vp_{\alpha'}}
 \Big)
\mr q_\alpha
\,,
$$
and hence $\mr q$ belongs to the hereditary $C^\ast$-sublgebra of $\mf F$ generated by $\mf C$. 
\qeb

\ebm

\section{Acknowledgements}

The authors are grateful to Hendrik Grundling for helpful discussions. One of us (M.S.) acknowledges funding by DFG under the grant SCHM 1652/2.

%%%%%%%%%%%%%%%%%%%%%%%%%%%%%%%%%%%%%%%%%%%%%%%%%%%%%%%%%%%%%%%%%%%%%%%%%%
%%%%%%%%%%%%%%%%%%%%%%%%%%%%%%%%%%%%%%%%%%%%%%%%%%%%%%%%%%%%%%%%%%%%%%%%%%


\begin{thebibliography}{99}

%%%%%%%%%%%%%%%%%%%%%%%%%%%%%%%%%%%%%%%%%%%%%%%%%%%%%%%%%%%%%%%%%%%%%%%%%%%%%%%%
%%%%%%%%%%%%%%%%%%%%%%%%%%%%%%%%%%%%%%%%%%%%%%%%%%%%%%%%%%%%%%%%%%%%%%%%%%%%%%%%%

\bibitem{AbrahamMarsden}
R.\ Abraham, J.E.\ Marsden: 
{\em Foundations of Mechanics.}
Benjamin/Cummings 1978

\bibitem{Asorey1}
M.\ Asorey, F.\ Falceto, J.\ Lopez, G. Luzon: Nodes, monopoles and confinement in $2+1$-dimensional Yang-Mills theory with the Chern-Simons term, Phys. Lett. B {\bf 153} (1985) 125-130

\bibitem{Asorey2}
M.\ Asorey: Maximal non-Abelian gauges and topology of the gauge orbit space, Nucl. Phys. B {\bf 551} (1999) 399-424

\bibitem{Blackadar}
B.\ Blackadar: {\em Operator algebras.} Springer, Heidelberg 2006

\bibitem{cfg}
 S.\ Charzy\'nski, J.\ Kijowski, G.\ Rudolph, M.\ Schmidt:
 On the stratified classical configuration space of lattice QCD.
 J.\ Geom.\ Phys.\ {\bf 55} (2005) 137--178

\bibitem{cfgtop}
 S.\ Charzy\'nski, G.\ Rudolph, M.\ Schmidt:
 On the topological structure of the stratified classical configuration space of lattice QCD. 
 J.\ Geom.\ Phys. {\bf 58} (2008) 1607--1623

\bibitem{DR}
T.\ Diez, G.\ Rudolph: Singular symplectic cotangent bundle reduction of gauge field theory, 
arXiv: 1812.04707v1 [math-ph]

\bibitem{emmroeme}
 C.\ Emmrich and H.\ Roemer:
 Orbifolds as configuration spaces of systems with gauge symmetries.
 Commun.\ Math.\ Phys.\  {\bf 129}, 69--94 (1990)

\bibitem{FRS}
E.\ Fischer, G.\ Rudolph, M.\ Schmidt:
A lattice gauge model of singular Marsden-Weinstein reduction. Part I. Kinematics. 
J.\ Geom.\ Phys.\ {\bf 57} (2007) 1193--1213

\bibitem{FJRS}
E.\ Fuchs, P.\ D.\ Jarvis, G.\ Rudolph, and M.\ Schmidt:
The Hilbert space costratification for the orbit type strata of SU(2)-lattice
gauge theory, J. Math. Phys.  {\bf 59}, 083505 (2018) 32p.

\bibitem{FSS}
K.\ Fukaya, P.\ Seidel, I.\ Smith: Exact Lagrangian submanifolds in simply-connected cotangent bundles. Invent.\ Math.\ {\bf 172} (2008) 1-–27

\bibitem{FuRS}
F.\ F\"urstenberg, G.\ Rudolph, M.\ Schmidt: Defining relations for the orbit type strata of $\SU(2)$-lattice gauge models. 
J.\ Geom.\ Phys.\ {\bf 119} (2017) 66-81  

\bibitem{GH1}
H.\ Grundling, C.\ A.\ Hurst: Algebraic Quantization
of Systems with a Gauge Degeneracy, Commun. Math. Phys. {\bf 98} (1985),
369–390.

\bibitem{GH2}
H.\ Grundling,  C.\ A.\ Hurst: The quantum theory of second class constraints:
kinematics, Commun. Math. Phys. {\bf 119} (1988), 75–93. 

\bibitem{GL}
H.\ Grundling, F.\ Lledo: Local Quantum Constraints, Rev. Math. Phys. {\bf 12}, No. 09 (2000)1159-1218

\bibitem{GR1}
H.\ Grundling, G.\ Rudolph: QCD on an Infinite Lattice, Commun. Math. Phys. {\bf 318} (2013)717–766

\bibitem{GR2}
 H.\ Grundling, G.\ Rudolph:  
 Dynamics for QCD on an infinite lattice. Commun. Math. Phys. {\bf 349} (2017) 1163-1202 
 
\bibitem{GuiSte}
V.\ Guillemin and S.\ Sternberg: 
Geometric quantization and multiplicities of group representations.
Invent.\ Math.\ {\bf 67} (1982) 515-–538

\bibitem{Hall:SBT}
 B.C.\ Hall: 
 The Segal-Bargmann ''coherent state'' transform for compact Lie groups.
 J.\ Funct.\ Anal.\ {\bf 122} (1994) 103--151

\bibitem{Hall:cptype}
 B.C.\ Hall: 
 Geometric quantization and the generalized Segal-Bargmann transform
 for Lie groups of compact type.
 Commun.\ Math.\ Phys.\ \textbf{226} (2002) 233--268

\bibitem{HallMitchell}
B.C.\ Hall, J.J. Mitchell: The Segal-Bargmann transform for noncompact symmetric spaces of the complex type. J.\ Funct.\ An.\ {\bf 227} (2005) 338--371

\bibitem{Scheck}
A.\ Heil, A.\ Kersch, N.\ A.\ Papadopolous, N.A. Reifenh\"auser, F.\ Scheck: Anomalies from nonfree action of the gauge group, Ann. Phys. {\bf 200} (1990) 206-215

\bibitem{HeinznerLoose}
 P.\ Heinzner, F.\ Loose:
 Reduction of complex Hamiltonian $G$-spaces.
 Geom.\ Funct.\ Anal.\ {\bf 4} (1994) 288--297

\bibitem{HoRS}
 M.\ Hofmann, G.\ Rudolph, M.\ Schmidt: 
 Orbit type stratification of the adjoint quotient of a compact semisimple Lie group.
 J.\ Math.\ Phys.\ {\bf 54} (2013) 083505

\bibitem{Hue:Quantization}
 J.\ Huebschmann:
 K\"ahler quantization and reduction.
 J.\ reine angew.\ Math.\ \textbf{591} (2006) 75--109


\bibitem{HRS}
 J.\ Huebschmann, G.\ Rudolph, M.\ Schmidt: 
 A lattice gauge model for quantum mechanics on a stratified space.
 Commun.\ Math.\ Phys. {\bf 286} (2009) 459--494
 
\bibitem{qcd1}
 P.D.\ Jarvis, J.\ Kijowski, G.\ Rudolph:
 On the structure of the observable algebra of QCD on the lattice.
 J.\ Phys.\ A {\bf 38} (2005) 5359--5377
 
\bibitem{KempfNess}
G.\ Kempf and L.\ Ness: The length of vectors in representation spaces.
Lecture Notes Math.\ {\bf 732}. Springer 1978, pp.\ 233--242

\bibitem{Kirwan}
F.\ Kirwan: Momentum maps and reduction in algebraic geometry.
Differential Geometry and its Applications 9 (1998) 135--171

\bibitem{qcd2}
 J.\ Kijowski, G.\ Rudolph:
 On the Gauss law and global charge for quantum chromodynamics.
 J.\ Math.\ Phys.\ {\bf 43} (2002) 1796--1808

\bibitem{qcd3}
 J.\ Kijowski, G.\ Rudolph:
 Charge superselection sectors for QCD on the lattice.
 J.\ Math.\ Phys.\ {\bf 46} (2005) 032303

\bibitem{KS}
J.\  Kogut,  L.\ Susskind:
 Hamiltonian formulation of Wilson's lattice gauge theories.
 Phys. Rev. D {\bf 11}  (1975) 395--408

\bibitem{Krantz}
S.G.\ Krantz: {\em Function Theory of Several Complex Variables}. AMS Chelsea Publishing, 2nd ed.\ 2001

\bibitem{Mumford}
 D.\ Mumford, J.\ Fogarty, F.\ Kirwan:
 {\em Geometric Invariant Theory.}
 Springer 1994

\bibitem{Murphy}
G.\ J.\ Murphy: {\em  $ C^*$-algebras and Operator Theory.} Boston, London, and San
Diego et.al.: Academic Press Limited, 1990

\bibitem{Mykytyuk}
I.V.\ Mykytyuk: Invariant K\"ahler structures on the cotangent bundles of compact symmetric spaces. Nagoya Math.\ J.\ {\bf 169} (2003) 191--217

\bibitem{OrtegaRatiu}
 J.-P.\ Ortega, T.S.\ Ratiu: 
 {\em Momentum Maps and Hamiltonian Reduction.}
 Progress in Mathematics, Vol.\ 222, Birkh\"auser 2004

\bibitem{Pedersen}
G.\ K.\ Pedersen: {\em $C^*$-algebras and their Automorphism Groups.} London Mathematical
Society. Monographs. London, San Diego, and New York et al.: Academic
Press Limited, 1989

\bibitem{Perutz}
T.\ Perutz, contribution to mathoverflow.net/questions/26776/k\"ahler-structure-on-cotangent-bundle, 16 December 2019

\bibitem{PflaumBook}
  M.J.\ Pflaum, \emph{Analytic and geometric study of stratified spaces}, Lecture
  Notes in Mathematics, vol. 1768, Springer-Verlag, Berlin, 2001.

\bibitem{RS}
 G.\ Rudolph, M.\ Schmidt: 
 On the algebra of quantum observables for a certain gauge model. 
 J.\ Math.\ Phys. {\bf 50} (2009) 052102
 
\bibitem{Buch}
 G.\ Rudolph, M.\ Schmidt: 
 {\it Differential Geometry and Mathematical Physics. Part I. Manifolds, Lie Groups and Hamiltonian Systems.}
 Springer 2013

\bibitem{RS-Buch2}
G.\ Rudolph, M.\ Schmidt: {\em Differential Geometry and Mathematical
Physics. Part II. Fibre Bundles, Topology and Gauge Fields.} Leipzig, Dodrecht:
Springer, 2017.

\bibitem{RSV}
G.\ Rudolph, M.\ Schmidt, I.P. Volobuev: Classification of gauge orbit types for $\SU(n)$-gauge theories, Mathematical Physics, Analysis and 
Geometry, {\bf 5} (2002) 201-241

\bibitem{Schlichenmaier}
M.\ Schlichenmaier: Singular projective varieties and quantization. In: Landsman N.P., Pflaum M., Schlichenmaier M. (eds) Quantization of Singular Symplectic Quotients. Progress in Mathematics, vol 198. Birkhäuser, Basel (2001)

\bibitem{Shan}
S.\ Shanmugadhasan: Canonical formalism of degenerate Lagrangians, J. Math Phys. 
{\bf 14} (1975) 677-687

\bibitem{SjamaarLerman}
 R.\ Sjamaar, E.\ Lerman:
 Stratified symplectic spaces and reduction.
 Ann.\ of Math.\ {\bf 134} (1991) 375--422

\bibitem{Stenzel}
M.B.\ Stenzel: The Segal-Bargmann transform on a symmetric space of compact type. 
J.\ Funct.\ Anal.\ {\bf 165} (1999) 44-–58

\bibitem{Szoeke}
R.\ Sz\"oke: Complex structures on tangent bundles of Riemannian manifolds. Math.\ Ann.\ {\bf 291} (1991) 409--428

\bibitem{Woodward}
C.\ Woodward: Moment maps and geometric invariant theory. Les cours du CIRM 1.1 (2010): 55-98. <http://eudml.org/doc/116365>

\end{thebibliography}
\end{document}